\documentclass[preprint]{aastex}
\pdfoutput=1

\usepackage{color}
\usepackage{url}
\usepackage{graphicx}
\usepackage{fancyhdr}
\graphicspath{{./}{figures/}}

\usepackage{listings}
\definecolor{lbcolor}{rgb}{0.9,0.9,0.9}
\usepackage{amsmath}
\DeclareMathOperator{\sinc}{sinc}
\DeclareMathOperator{\III}{III}

\newcommand{\Fig}[1]{Figure~\ref{fig:#1}}
\newcommand{\fig}[1]{Figure~\ref{fig:#1}}
\newcommand{\figs}[2]{Figures~\ref{fig:#1}-\ref{fig:#2}}
\newcommand{\figlabel}[1]{\label{fig:#1}}
\newcommand{\Eq}[1]{Equation~\ref{eq:#1}}
\newcommand{\eq}[1]{\Eq{#1}}
\newcommand{\eqs}[2]{Equations~\ref{eq:#1}-\ref{eq:#2}}
\newcommand{\eqlabel}[1]{\label{eq:#1}}
\newcommand{\Sect}[1]{Section~\ref{sect:#1}}
\newcommand{\sect}[1]{\Sect{#1}}

\newcommand{\sectlabel}[1]{\label{sect:#1}}

\usepackage[normalem]{ulem}

\begin{document}

\title{Understanding the Lomb-Scargle Periodogram}

%\title{Understanding the Lomb-Scargle Periodogram\\
%       {\it (Draft Version \today)}}

% Add ``Draft version'' footer
%\pagestyle{fancy}
%\fancyhf{}\renewcommand{\headrulewidth}{0pt} % remove header titles
%\cfoot{\it \footnotesize $\sim$ Draft Version, \today $\ \sim$}

\newcommand{\escience}{1}
\author{Jacob T. VanderPlas\altaffilmark{\escience}}
\altaffiltext{\escience}{eScience Institute, University of Washington}
%\author{Jacob T. VanderPlas}
%\affil{eScience Institute, University of Washington}

\begin{abstract}
The Lomb-Scargle periodogram is a well-known algorithm for detecting and
characterizing periodic signals in unevenly-sampled data.
This paper presents a conceptual introduction to the Lomb-Scargle periodogram
and important practical considerations for its use.
Rather than a rigorous mathematical treatment, the goal of this paper is to
build intuition about what assumptions are implicit in the use of the
Lomb-Scargle periodogram and related estimators of periodicity,
so as to motivate important practical considerations required
in its proper application and interpretation.
\end{abstract}


\section{Introduction}
\sectlabel{introduction}

The Lomb-Scargle periodogram \citep{Lomb76, Scargle82}
is a well-known algorithm for detecting and characterizing
periodicity in unevenly-sampled
time-series, and has seen particularly wide use within the astronomy community.
As an example of a typical application of this method,
consider the data shown in \fig{LINEAR-data}:
this is an irregularly-sampled timeseries showing a single object from the
LINEAR survey \citep{LINEAR1, LINEAR3}, with un-filtered magnitude measured
280 times over the course of five and a half years.
By eye, it is clear that the brightness of the object varies in time with a
range spanning approximately 0.8 magnitudes, but what is not immediately clear
is that this variation is periodic in time.
The Lomb-Scargle periodogram is a method that allows efficient computation of
a Fourier-like power spectrum estimator from such unevenly-sampled data,
resulting in an intuitive means of determining the period of oscillation.

\begin{figure}[ht]
\centering
\includegraphics[width=\textwidth]{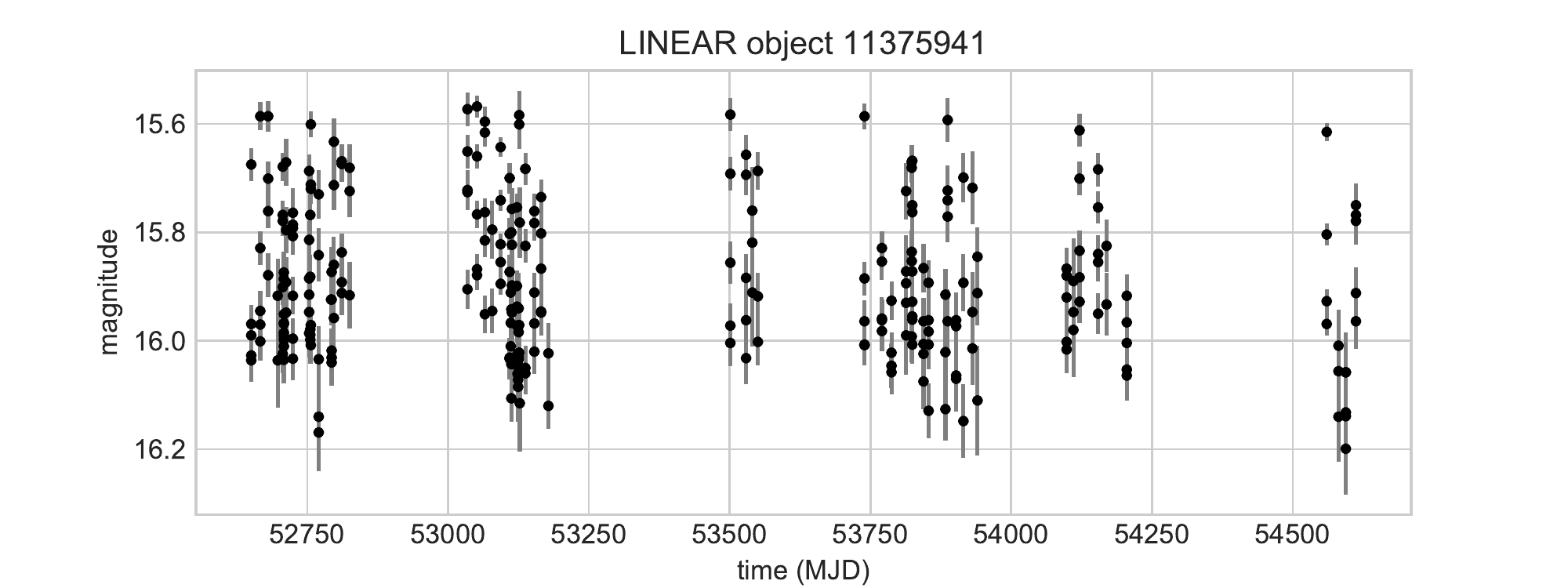}
\caption{Observed light curve from LINEAR object ID 11375941. Uncertainties
  are indicated by the gray errorbars on each point.
  Python code to reproduce this figure, as well as all other figures
  in this manuscript, is available at {\tt http://github.com/jakevdp/PracticalLombScargle/}
  \figlabel{LINEAR-data}
}
\end{figure}

\begin{figure}[ht]
\centering
\includegraphics[width=\textwidth]{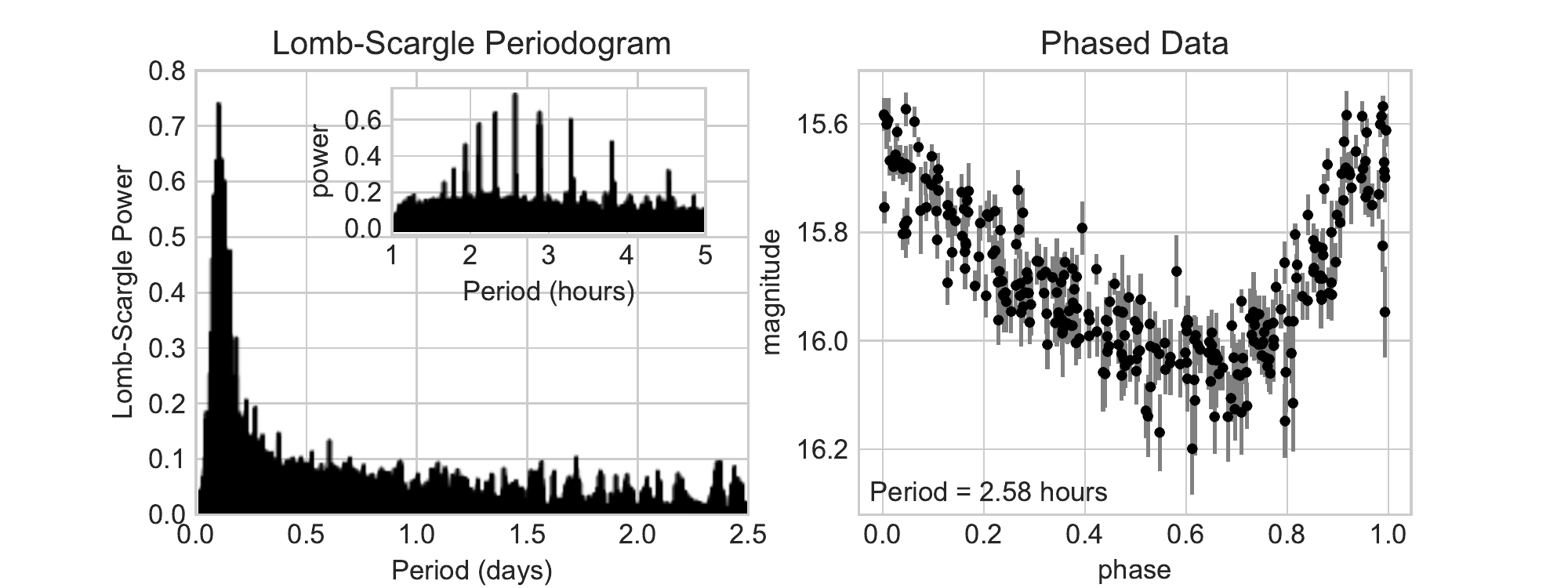}
\caption{{\it Left panel:} the Lomb-Scargle periodogram computed from the data
    shown in \fig{LINEAR-data}, with an inset detailing the region near the peak.
    {\it Right panel:} the input data in \fig{LINEAR-data}, folded over the
    detected 2.58-hour period to show the coherent periodic variability.
    For more discussion of this particular object, see \citep{LINEAR3}.
    \figlabel{LINEAR-power}
}
\end{figure}

The Lomb-Scargle periodogram computed from this data is shown in the left
panel  of \fig{LINEAR-power}.
The Lomb-Scargle periodogram here yields an estimate of the Fourier
power as a function of period of oscillation, from
which we can read-off the period of oscillation of approximately 2.58 hours.
The right panel of \fig{LINEAR-power} shows a folded visualization of
the same data as \fig{LINEAR-data} -- i.e.{} plotted as a function of phase
rather than time.

Often this is exactly how the Lomb-Scargle periodogram is presented: as a
clean, well-defined procedure to generate a power spectrum and to
detect the periodic component in an unevenly-sampled dataset.
In practice, however, there are a number of subtle issues that must be
considered when applying a Lomb-Scargle analysis to real-world datasets.
Here are a few questions in particular that we might wish to ask about
the results in \fig{LINEAR-power}:
\begin{enumerate}
  \item How does the Lomb-Scargle periodogram relate to the classical
    Fourier power spectrum?
  \item What is the source of the pattern of multiple peaks revealed by the
    Lomb-Scargle Periodogram?
  \item What is the largest frequency (i.e. Nyquist-like limit) that such
    an analysis is sensitive to?
  \item How should we choose the spacing of the frequency grid for our
    periodogram?
  \item What assumptions, if any, does the Lomb-Scargle approach make
    regarding the form of the unknown signal?
  \item How should we understand and report the uncertainty of the
    determined frequency?
\end{enumerate}
Quantitative treatments of these sorts of questions are presented in various
textbooks and review papers, but I have not come across any single concise
reference that gives a good intuition for how to think about such questions.
This paper seeks to fill that gap, and provide a practical,
just-technical-enough guide to the effective use of the Lomb-Scargle method
for detection and characterization of periodic signals.
This paper does not seek a complete or rigorous treatment of the mathematics
involved, but rather seeks to develop the intuition of {\it how} to think
about these questions, with references to relevant technical treatments.

\subsection{Why Lomb-Scargle?}
\sectlabel{why-lomb-scargle}

Before we begin exploring the Lomb-Scargle periodogram in more depth,
it is worth briefly considering the broader context of methods for
detecting and characterizing periodicity in time series.
First, it is important to note that there are many different modes of
time series observation.
Point observation like that shown in \fig{LINEAR-data} is typical of
optical astronomy: values (often with uncertainties) are measured at
discrete point in time, which may be equally or unequally-spaced.
Other modes of observation---e.g., time-tag-events, binned event data,
time-to-spill events, etc.---are common in high-energy astronomy and other
areas. We will not consider such event-driven data modes here, but note
that there have been some interesting explorations of unified statistical
treatments of all of the above modes of observation
\citep[e.g.][]{Scargle98, Scargle2002}.

Even limiting our focus to point observations, there are a large number of
complementary techniques for periodic analysis, which generally can be
categorized into a few broad categories:
\begin{description}
  \item[Fourier Methods] are based on the Fourier transform, power spectra,
    and closely related correlation functions. These methods include the
    classical or Schuster periodogram \citep{Schuster98},
    the Lomb-Scargle periodogram
    \citep{Lomb76, Scargle82}, the correlation-based method of
    \citet{Edelson88}, and related approaches
    \citep[see also][for a discussion of wavelet transforms in this
      context]{Foster96}.
  \item[Phase-folding Methods] depend on folding
    observations as a function of phase, computing a cost function across
    the phased data (often within bins constructed across phase space)
    and optimizing this cost function across candidate frequencies.
    Some examples are
    String Length \citep{Dworetsky83},
    Analysis of Variance \citep{Schwarzenberg-Czerny89},
    Phase Dispersion Minimization \citep{Stellingwerf78},
    the Gregory-Laredo method \citep{Gregory92},
    and the conditional entropy method \citep{Graham2013}.
    Methods based on correntropy are similar in spirit, but do not always
    require explicit phase folding \citep{Huijse2011,Huijse2012}.
  \item[Least Squares Methods] involve fitting a model to the data at each
    candidate frequency, and selecting the frequency which maximizes the likelihood.
    The Lomb-Scargle periodogram also falls in this category
    (see \sect{schuster-to-lomb-scargle}), as does the Supersmoother approach
    \citep{Reimann94}. Other studies recommend statistics other than
    least square residuals; see, e.g., the orthogonal polynomial fits of
    \citet{Schwarzenberg-Czerny96}.
  \item[Bayesian Approaches] apply Bayesian probability theory to the problem,
    often in a similar manner to the phase-folding and/or least-squares
    approaches.
    Examples are the generalized Lomb-Scargle models of \citet{Bretthorst88},
    the phase-binning model of \citet{Gregory92}, Gaussian process models
    \citep[e.g.][]{Wang2012}, and models based on stochastic processes
    \citep[e.g.][]{Kelly14}.
\end{description}
Various reviews have been undertaken to compare the efficiency and effectiveness
of the available methods; for example, \citet{Schwarzenberg-Czerny99}
focuses on the statistical properties of methods and recommends those
based on smooth model fits over methods based on phase binning,
while \citet{Graham2013b} instead take an empirical approach and find that
when considering detection efficiency in real datasets, no suitably efficient
algorithm universally outperforms the others.

In light of this breadth of available methods, why limit our focus here to the
Lomb-Scargle periodogram?
One reason is cultural: the Lomb-Scargle periodogram is perhaps the best-known
technique to compute periodicity of unequally-spaced data in astronomy and
other fields, and so is the first tool many will reach for
when searching for periodic content in a signal.
But there is a deeper reason as well; it turns out that Lomb-Scargle occupies
a unique niche: it is motivated by Fourier analysis (see \sect{schuster-to-lomb-scargle}),
but it can also be viewed as a least squares method
(see \sect{lomb-scargle-extensions}).
It can be derived from the principles
of Bayesian probability theory (see \sect{bayesian-periodograms}), and has
been shown to be closely related to bin-based phase-folding techniques under
some circumstances \citep[see][]{Swingler89}.
Thus, the Lomb-Scargle periodogram occupies a unique point of correspondence
between many classes of methods, and so provides a focus for discussion that
has bearing on considerations involved in {\it all} of these methods.

\subsection{Outline}
The remainder of this paper is organized as follows:\\
{\bf\sect{continuous-fourier-transform}} presents a review of the
continous Fourier transform and some of its useful properties,
including defining the notion of a power spectrum (i.e. classical/Schuster
periodogram) for detecting periodic content in a signal.\\
{\bf\sect{window-functions}} builds on these properties to explore how
different observation patterns (i.e. window functions) affect the
periodogram, and discusses a conceptual understanding of the
Nyquist limiting frequency.\\
{\bf\sect{non-uniform-sampling}} considers non-uniformly sampled signals as
a special case of an observational window, and shows that the Nyquist-like
limit for this case is quite different than what many practitioners in
the field often assume.\\
{\bf\sect{schuster-to-lomb-scargle}} introduces the Lomb-Scargle periodogram,
a modified version of the classical periodogram for unevenly-sampled data,
as well as the motivation behind these modifications.\\
{\bf\sect{lomb-scargle-extensions}} discusses a complementary view of the
Lomb-Scargle periodogram as the result of least-squares model fitting,
as well as some extensions enabled by that viewpoint.\\
{\bf\sect{practical-considerations}} builds on the concepts introduced
in earlier sections to discuss important practical considerations for
the use of the Lomb-Scargle periodogram, including the choice of
frequency grid, uncertainties and false alarm probabilities, and modes of
failure that should be accounted for when working with real-world data.\\
{\bf\sect{conclusion}} concludes and summarizes the key recommendations
for practical use of the Lomb-Scargle periodogram.

\section{Background: the Continuous Fourier Transform}
\sectlabel{continuous-fourier-transform}

In order to understand how we should interpret the Lomb-Scargle periodogram, we will first briefly step back and review the subject of Fourier analysis of continuous signals.
Consider a continuous signal $g(t)$.
Its Fourier transform is given by the following integral, where $i\equiv\sqrt{-1}$ denotes the imaginary unit:
\begin{equation}
    \hat{g}(f) \equiv \int_{-\infty}^\infty g(t) e^{-2\pi i f t} dt
    \eqlabel{FT-def}
\end{equation}
The inverse relationship is given by:
\begin{equation}
    g(t) \equiv \int_{-\infty}^\infty \hat{g}(f) e^{+2\pi i f t} df
    \eqlabel{IFT-def}
\end{equation}
For convenience we will also define the Fourier transform operator
$\mathcal{F}$, such that
\begin{eqnarray}
    \mathcal{F}\{g\} &=& \hat{g} \\
    \mathcal{F}^{-1}\{\hat{g}\} &=& g
\end{eqnarray}
The functions $g$ and $\hat{g}$ are known as a {\it Fourier pair}, which we
will sometimes denote as $g \Longleftrightarrow \hat{g}$.

\subsection{Properties of the Fourier Transform}
\sectlabel{ft-properties}
The continuous Fourier transform has a number of useful properties that we will make use of in our discussion.

\begin{description}
   \item[The Fourier transform is a linear operation.]
     That is, for any constant $A$ and any functions $f(t)$ and $g(t)$,
     we can write:
     \begin{eqnarray}
       \mathcal{F}\{f(t) + g(t)\} &=& \mathcal{F}\{f(t)\} + \mathcal{F}\{g(t)\}\nonumber\\
       \mathcal{F}\{A f(t)\} &=& A\mathcal{F}\{f(t)\}
     \end{eqnarray}
     Both identities follow from the linearity of the Fourier integral.

   \item[The Fourier transform of sinusoid with frequency $f_0$ is a sum of delta functions at $\pm f_0$.]
     From the integral definition of the Dirac delta function\footnote{
       $\delta(f)\equiv\int_{-\infty}^\infty e^{-2\pi i x f}df$},
     we can write
     \begin{equation}
       \mathcal{F}\{e^{2\pi f_0 t}\} = \delta(f - f_0).
       \eqlabel{delta-FT}
     \end{equation}
     Using Euler's formula\footnote{$e^{ix} = \cos x + i\sin x$}
     for the complex exponential along with the linearity
     of the Fourier transform leads to the following identities:
     \begin{eqnarray}
       \mathcal{F}\{\cos(2\pi f_0 t)\} &=& \frac{1}{2}\left[\delta(f - f_0) + \delta(f + f_0)\right]\nonumber\\
       \mathcal{F}\{\sin(2\pi f_0 t)\} &=& \frac{1}{2i}\left[\delta(f - f_0) - \delta(f + f_0)\right].
     \end{eqnarray}
     In other words, a sinusoidal signal with frequency $f_0$ has a Fourier
     transform consisting of a weighted sum of delta functions at $\pm f_0$.

   \item[A time-shift imparts a phase in the Fourier transform.]
     Given a well-behaved function $g(t)$ we can use a transformation of
     variables to derive the following identity:
     \begin{equation}
       \mathcal{F}\{g(t - t_0)\} = \mathcal{F}\{g(t)\} e^{-2\pi i ft_0}
     \end{equation}
     Notice that the time-shift does not change the
     amplitude of the resulting transform, but only the phase.
\end{description}
These properties taken together make the Fourier transform quite useful for the study of periodic signals.
The linearity of the transform means that any signal made up of a sum of
sinusoidal components will have a Fourier transform consisting of a sum
of delta functions marking the frequencies of those sinusoids: that is,
the Fourier transform directly measures periodic content in a continuous function.

Further, if we compute the squared amplitude of the resulting transform, we
can both do away with complex components and remove the phase imparted by
the choice of temporal baseline; this squared amplitude is usually known as the
{\it power spectral density} or simply the {\it power spectrum}:
\begin{equation}
  \mathcal{P}_g \equiv \left|\mathcal{F}\{g\}\right|^2
  \eqlabel{power-spectrum}
\end{equation}
The power spectrum of a function is a positive real-valued function of the
frequency $f$ that quantifies the contribution of each frequency $f$ to
the total signal.
Note that if $g$ is real-valued, it follows that $P_g$ is an even function;
{i.e.} $\mathcal{P}_g(f) = \mathcal{P}_g(-f)$.

\subsection{Some Useful Fourier Pairs}

\begin{figure}[ht]
\centering
\includegraphics[width=\textwidth]{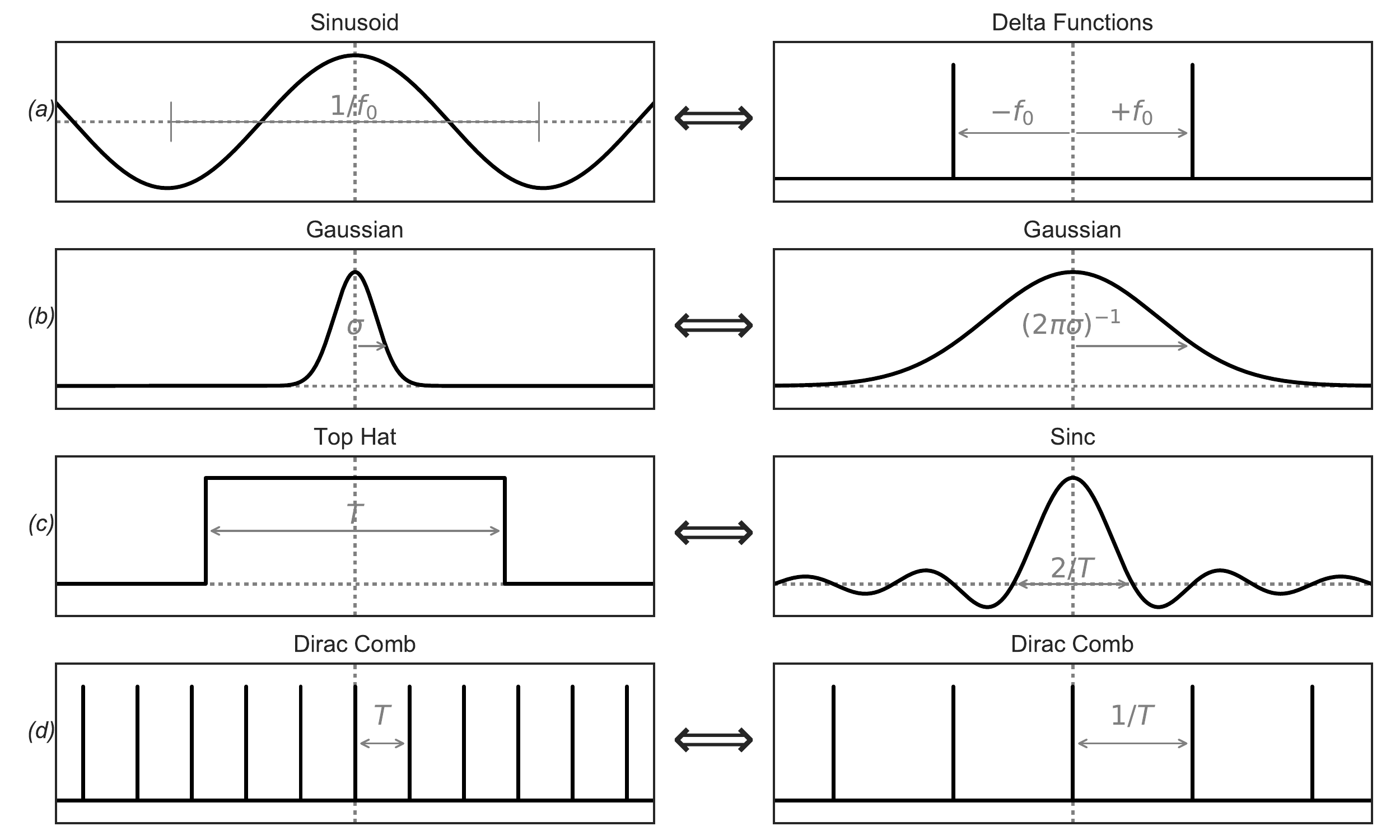}
\caption{Visualization of important Fourier pairs.\figlabel{fourier-pairs}}
\end{figure}

We have already discussed that the Fourier transform of a complex exponential is a single delta function.
This is just one of many Fourier pairs to keep in mind as we progress to understanding the Lomb-Scargle periodogram.
We list a few more important pairs here (see \Fig{fourier-pairs} for a visual representation of the following pairs):

\begin{description}
  \item[The Fourier transform of a sinusoid is a pair of Delta functions.]
    (See \fig{fourier-pairs}{\it a})
    \begin{equation}
      \mathcal{F}\{\cos(2\pi f_0 t)\} = \frac{1}{2}\left[\delta(f-f_0) + \delta(f+f_0)\right]
    \end{equation}
    We saw this above, but repeat it here for completeness.

   \item[The Fourier transform of a Gaussian is a Gaussian.]
    (See \fig{fourier-pairs}{\it b})
     \begin{equation}
       \mathcal{F}\{{\rm N}(t; \sigma)\} = \frac{1}{\sqrt{2\pi\sigma^2}}{\rm N}\left(f;\frac{1}{2\pi\sigma}\right)
     \end{equation}
     The Gaussian function ${\rm N}(t, \sigma)$ is given by
     \begin{equation}
       {\rm N}(t; \sigma) \equiv \frac{1}{\sqrt{2\pi\sigma^2}}e^{-t^2/(2\sigma^2)}
     \end{equation}

  \item[The Fourier transform of a rectangular function is a sinc function.]
    (See \fig{fourier-pairs}{\it c})
    \begin{equation}
      \mathcal{F}\{\Pi_T(t)\} = \sinc(f T)
    \end{equation}
    The rectangular function, $\Pi(t)$, is a normalized symmetric function that
    is uniform within a range given by $T$, and zero elsewhere:
    \begin{equation}
      \Pi_T(t)  \equiv \left\{
      \begin{array}{ll}
        1 / T, & |t| \le T / 2 \\
        0,     & |t| > T / 2
      \end{array}
      \right.
    \end{equation}
    The sinc function is given by the standard definition:
    \begin{equation}
      \sinc(x) \equiv \frac{\sin(\pi x)}{\pi x}
    \end{equation}

  \item[The Fourier transform of a Dirac comb is a Dirac comb.]
    (See \fig{fourier-pairs}{\it d})
    \begin{equation}
      \mathcal{F}\{\III_T(t)\} = \frac{1}{T}\III_{1/T}(f)
    \end{equation}
    The Dirac comb, $\III_T(t)$, is an infinite sequence of Dirac delta functions placed at even intervals of size $T$:
    \begin{equation}
      \III_T(t) \equiv \sum_{n=-\infty}^\infty \delta(t - nT)
      \eqlabel{dirac-comb}
    \end{equation}
\end{description}
Notice in each of these Fourier pairs the reciprocity
of scales between a function and its Fourier transform:
a narrow function will have a broad transform, and vice versa.
More quantitatively, a function with a characteristic scale $T$ will in
general have a Fourier transform with characteristic scale of $1/T$.
Such reciprocity is central to many diverse applications of Fourier analysis,
from electrodynamics to music theory to quantum mechanics.
For our purposes, this property will turn out to be quite
important as we push further in understanding the Lomb-Scargle periodogram.

\subsection{The Convolution Theorem}

A final property of the Fourier transform that we will discuss here is its
ability to convert convolutions into point-wise products.
A convolution of two functions, usually denoted by the $\ast$ symbol,
is defined as follows:
\begin{equation}
  [f \ast g](t) \equiv \int_{-\infty}^\infty f(\tau)g(t - \tau) d\tau
  \eqlabel{convolution-definition}
\end{equation}
From the definition, it is clear that a convolution amounts to ``sliding'' one
function past the other, integrating at each step.
Such an operation is commonly used, for example, in smoothing a function,
as visualized in \Fig{convolution} for a rectangular smoothing window.

\begin{figure}[ht]
  \centering
  \includegraphics[width=\textwidth]{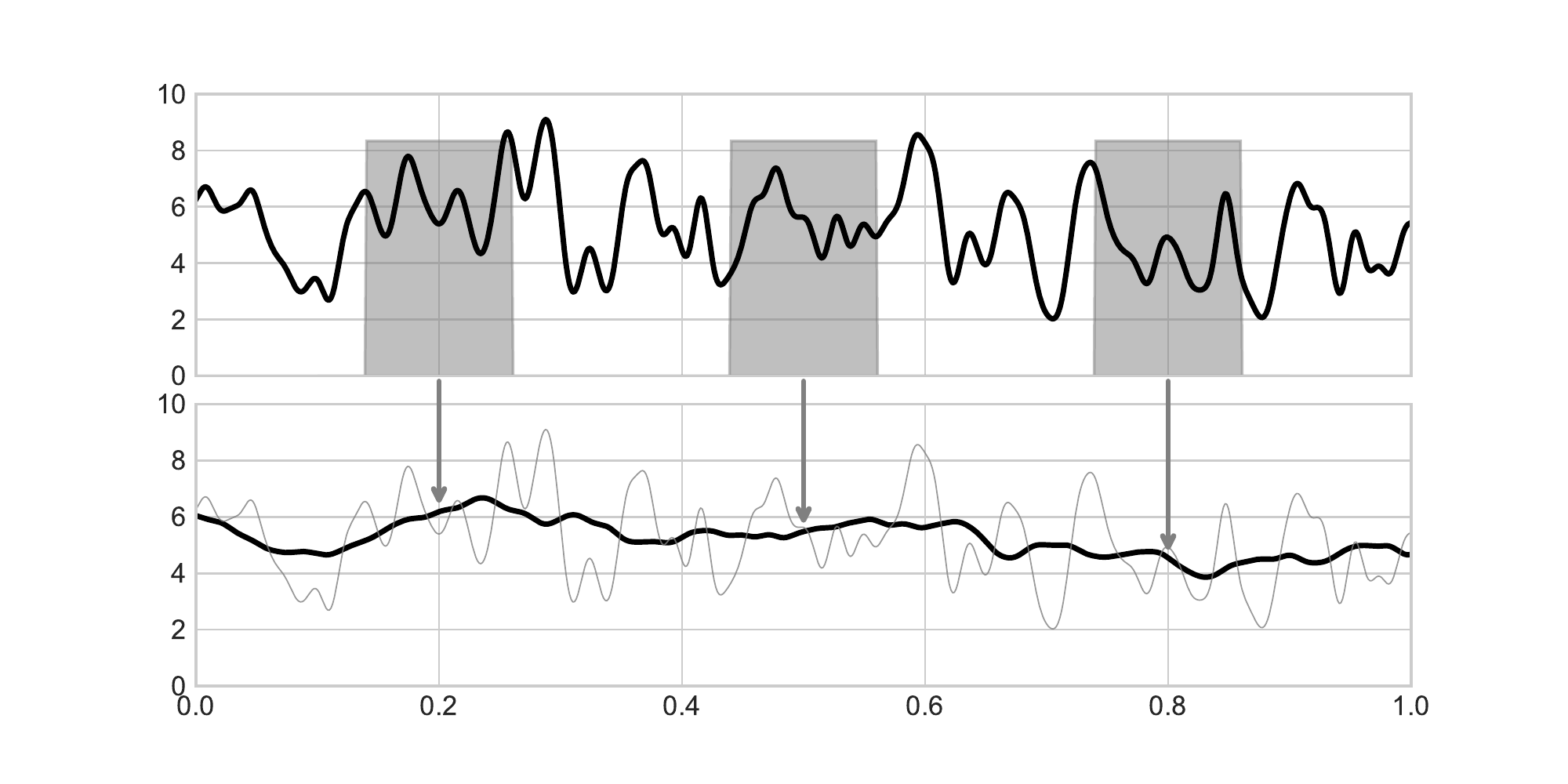}
  \caption{Visualization of a convolution between a continuous signal and a rectangular smoothing kernel.
    The normalized rectangular window function slides across the domain (upper panel),
    such that at each point the mean of the values within the window are
    used to compute the smoothed function (lower panel).
    \figlabel{convolution}}
\end{figure}

Given this definition of a convolution, it can be shown that the Fourier transform of a convolution is the point-wise product of the individual
Fourier transforms:
\begin{equation}
  \mathcal{F}\{f \ast g\} = \mathcal{F}\{f\} \cdot \mathcal{F}\{g\}
  \eqlabel{convolution-theorem}
\end{equation}
In practice, this can be a much more efficient means of computing a convolution
than to directly solve at each time $t$ the integral over $\tau$ that appears
in \eq{convolution-definition}.
The identity in \eq{convolution-theorem} is known as the
{\it convolution theorem}, and is illustrated in \fig{convolution-theorem}.
\begin{figure}[ht]
  \centering
  \includegraphics[width=\textwidth]{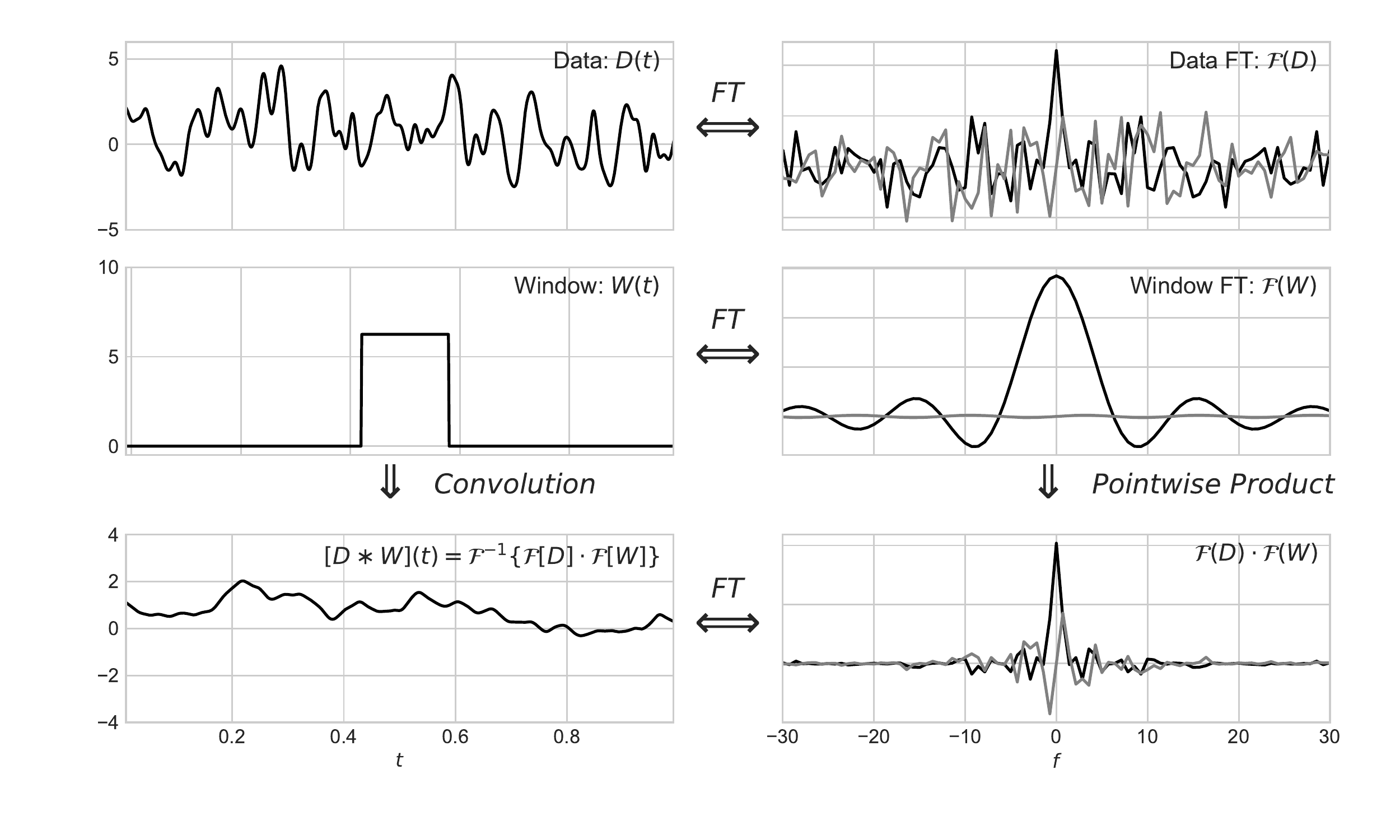}
  \caption{Visualization of the convolution theorem (\eq{convolution-theorem}).
    Recall that the Fourier transform of
    a convolution is the pointwise product of the two Fourier transforms.
    In the right panels, the black and gray lines represent the real and
    imaginary part of the transform, respectively.
    \figlabel{convolution-theorem}}
\end{figure}
An important corrollary is that the Fourier transform of a product is a convolution of the two transforms:
\begin{equation}
  \mathcal{F}\{f \cdot g\} = \mathcal{F}\{f\} \ast \mathcal{F}\{g\}
  \eqlabel{convolution-theorem-inverse}
\end{equation}
We will see that these properties of the Fourier transform become essential
when thinking about frequency components of time-domain measurements.

\section{Window Functions: From Idealized to Real-world signals}
\sectlabel{window-functions}

Until now we have been discussing Fourier transforms of continuous signals
which are well-defined for all times $-\infty < t < \infty$.
Real-world temporal measurements of a signal, however, only involve some
finite span of time, at some finite rate of sampling.
In either case, the resulting data can be described by a point-wise product
of the true underlying continuous signal with a window function describing
the observation.
For example, a continuous signal measured over a finite duration is described
by a rectangular window function spanning the duration of the observation,
and a signal measured at regular intervals is described by a Dirac comb window
function marking those measurement times.

The Fourier transform of measured data in these cases, then, is not the
transform of the continuous underlying function, but rather the transform
of the {\it point-wise product of the signal and the observing window.}
Symbolically, if the signal is $g(t)$ and the window is $W(t)$, the observed
function is
\begin{equation}
  g_{obs}(t) = g(t)W(t),
\end{equation}
and by the convolution theorem, its transform is a convolution
of the signal transform and the window transform:
\begin{equation}
  \mathcal{F}\{g_{obs}\} = \mathcal{F}\{g\} \ast \mathcal{F}\{W\}.
\end{equation}
This has some interesting consequences for the use and interpretation of
periodograms, as we shell see.

\subsection{Effect of a Rectangular Window}

\begin{figure}[ht]
  \centering
  \includegraphics[width=\textwidth]{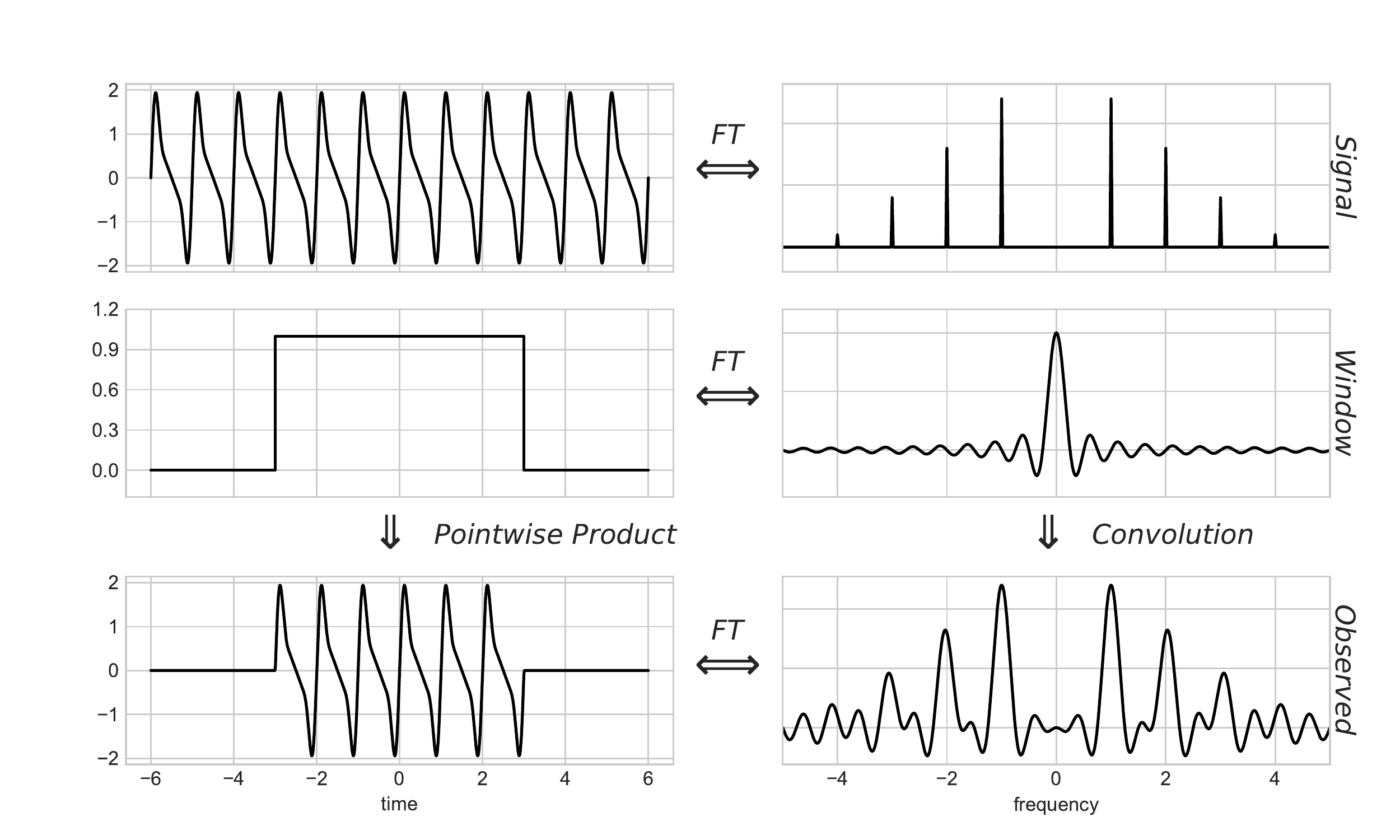}
  \caption{Visualization of the effect on the Fourier transform of a
    rectangular observing window (i.e., a continuous signal observed in its
    entirety within a finite range of time). The function used here is
    $g(t) = 1.2\sin(2\pi t) + 0.8\sin(4\pi t) + 0.4\sin(6\pi t) + 0.1\sin(8\pi t)$; The observed Fourier
    transform is a convolution of the true transform (here a series of Delta
    functions indicating the component frequencies) and the window transform
    (here a narrow sinc function).
    \figlabel{rectangular-window}}
\end{figure}

First, let's consider the case of observing a continuous periodic signal over
a limited span of time: \fig{rectangular-window} shows
a continuous periodic function observed only within the window $-3<t<3$.
The observed signal in this case can be understood as the pointwise product of
the underlying infinite periodic signal with a rectangular window function.
By the convolution theorem, the Fourier transform will be given by the
convolution of the transform of the underlying function (here a set of
delta functions at the component frequencies) and the transform of the window
function (here a sinc function).
For the purely periodic signal like the one seen in \fig{rectangular-window},
this convolution has the effect of replacing each delta function with a
sinc function.
Because of the inverse relationship between the width of the window and the
width of its transform (see \Fig{fourier-pairs}), it follows that a wider
observing window leads to proportionally less spread in the Fourier
transform of the observed function.

\subsection{The Dirac Comb and the Discrete Fourier Transform}

\begin{figure}[ht]
  \centering
  \includegraphics[width=\textwidth]{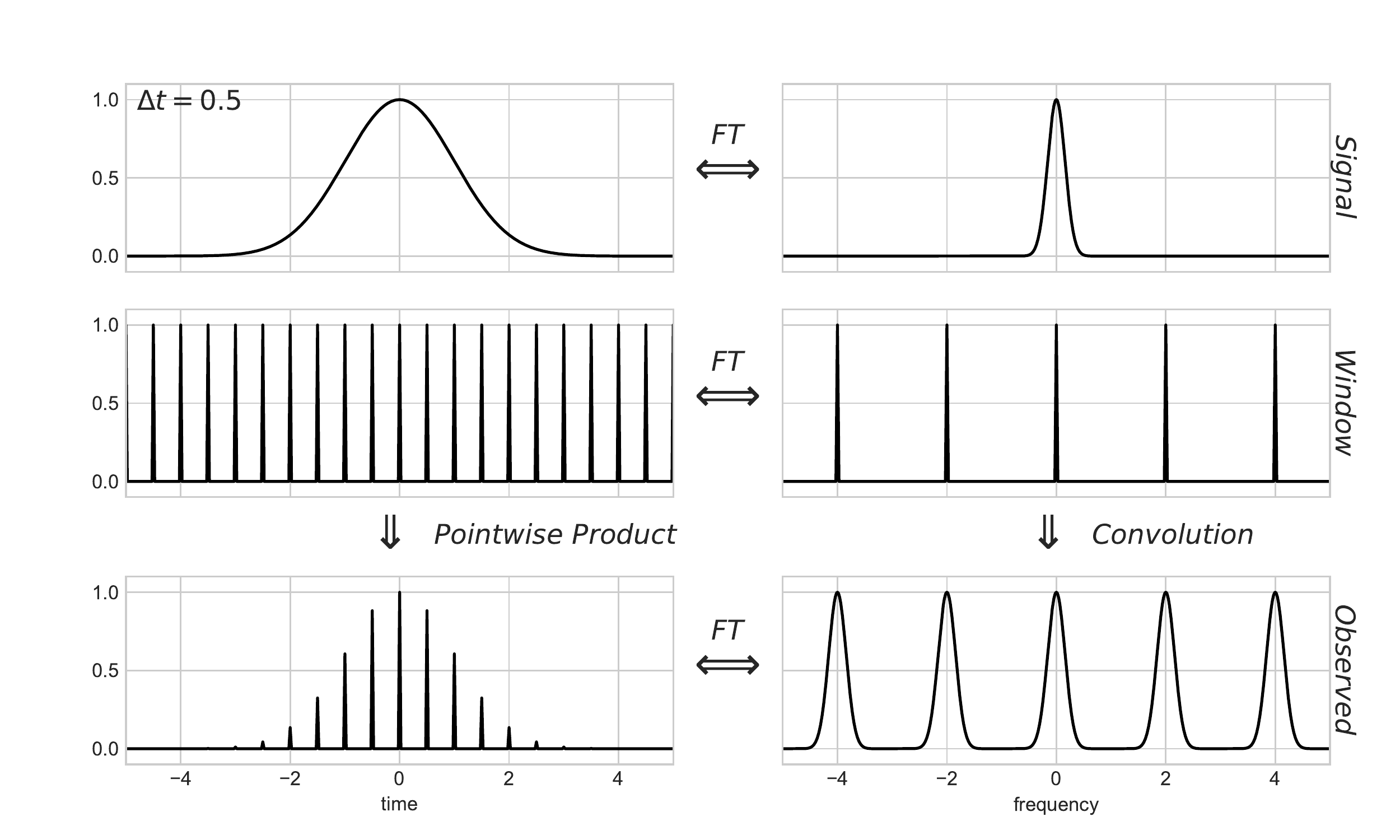}
  \caption{Visualization of the effect on the Fourier transform of a
    Dirac Comb observing window (i.e., a long string of evenly-spaced
    discrete observations). The observed Fourier
    transform is a convolution of the true transform (here a localized
    Gaussian) and the window transform (here another Dirac comb).
    \figlabel{comb-window-1}}
\end{figure}

Another window function that commonly arises is when a continuous signal is
sampled (nearly) instantaneously at regular intervals.
Such an observation can be thought of as a point-wise product between the true
underlying signal and a Dirac comb with the $T$ parameter matching the spacing
of the observations; this is illustrated in \fig{comb-window-1}.
Interestingly, because the Fourier transform of a Dirac comb is another Dirac
comb, the effect of such an observing window is to create a long sequence
of aliases of the underlying transform with a spacing of $1/T$.
With this in mind, we can be assured in this case that evaluating the
observed transform in the range $0 \le f < 1/T$ is sufficient to capture
all the available frequency information:
the signal outside that range is a sequence of identical aliases of
what lies within that range.

\subsubsection{The Nyquist Limit}
\sectlabel{nyquist}

The example in \fig{comb-window-1} is somewhat of a best-case scenario, because
the true Fourier transform values are non-zero only within a range of
width $1/T$.
If we increase the time between observations, decreasing the spacing of the
frequency comb, the true transform no longer ``fits'' inside the window
transform, and we will have a situation similar to that in \fig{comb-window-2}.
The result is a mixing of different portions of the signal, such that
the true Fourier transform {\it cannot be recovered}
from the transform of the observed data!

\begin{figure}[ht]
  \centering
  \includegraphics[width=\textwidth]{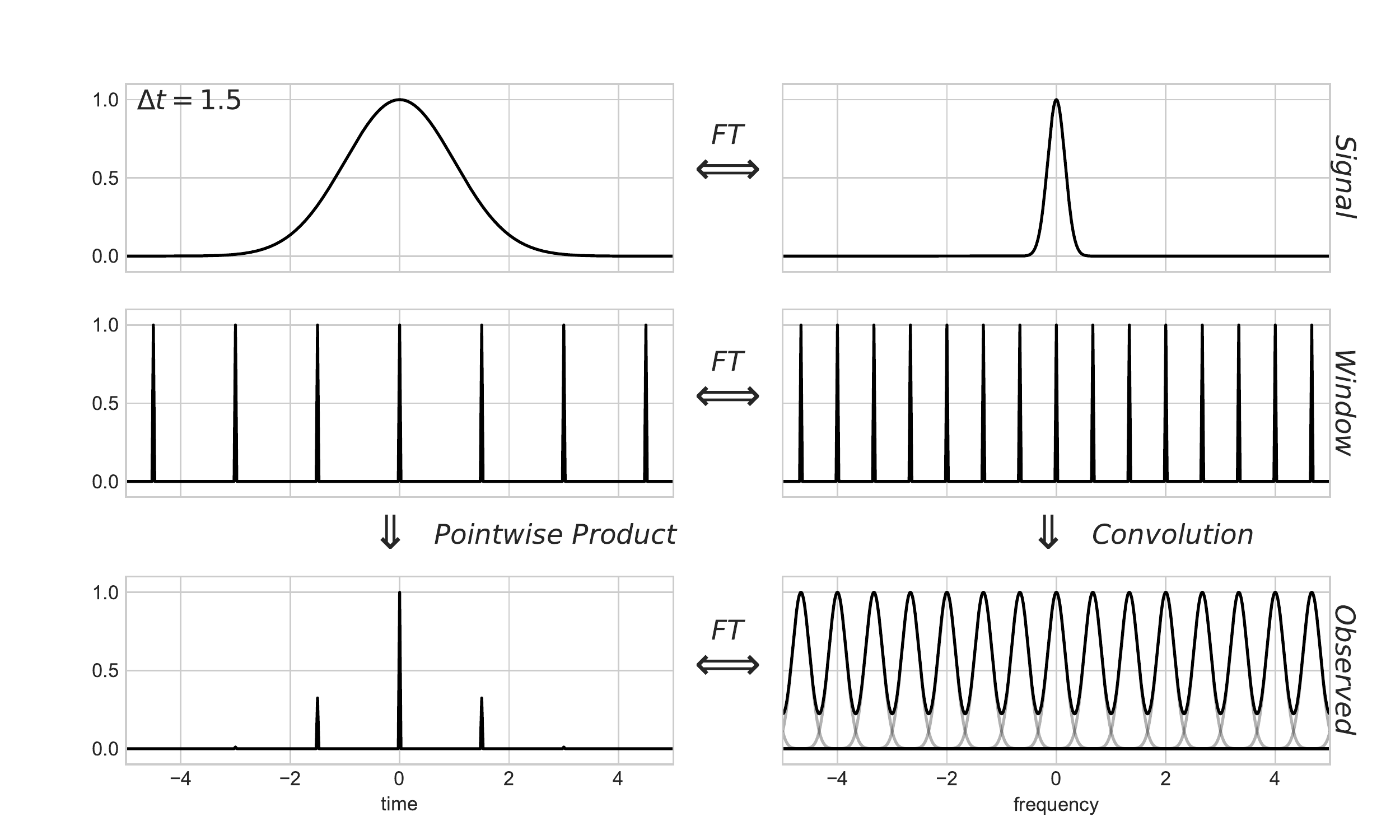}
  \caption{Repeating the visualization from \fig{comb-window-1}, but here with
    a lower sampling rate. The result is that the Fourier transform of the
    window function (middle right) has spacing narrower than the Fourier
    transform of the signal (upper right), meaning the observed Fourier
    transform (lower right) has
    aliasing of signals, such that not all frequency information can be
    recovered. This is the reason for the famous Nyquist sampling theorem,
    which conceptually says that only a function whose Fourier transform can
    fit entirely between the ``teeth'' of the comb is able to be fully
    recovered for regularly-spaced observations.
    \figlabel{comb-window-2}}
\end{figure}

This implies that if we have a regularly-sampled function with a sampling
rate of $f_0 = 1/T$, we can only fully recover the frequency information
if the signal is {\it band-limited} between frequencies $\pm f_0/2$.
This is one way to motivate the famous Nyquist sampling limit, which
approaches the question from the other direction and states
that to fully represent the frequency content of a ``band-limited signal''
whose Fourier transform is zero outside the range $\pm B$,
we must sample the data with a rate of at least $f_{Ny} = 2B$.

\subsubsection{The Discrete Fourier Transform}
When a continuous function is sampled at regular intervals, the delta functions
in the Dirac comb window serve to collapse the Fourier integral into a Fourier
sum, and in this manner we can arrive at the common form of the discrete Fourier
transform.
Suppose we have a true (infinitely long and continuous) signal $g(t)$, but
we observe it only at a regular grid with spacing $\Delta t$. In this case, our
observed signal is $g_{obs} = g(t) \III_{\Delta t}(t)$ and its Fourier transform is
\begin{equation}
  \hat{g}_{obs}(f) = \sum_{n=-\infty}^\infty g(n\Delta t) e^{-2\pi i f n \Delta t},
  \eqlabel{DFT-f-inf}
\end{equation}
which follows directly from \eq{FT-def} and \eq{dirac-comb}.

In the real world, however, we will not have an infinite number of observations,
but rather a finite number of samples $N$.
We can choose the coordinate system appropriately and define
$g_n \equiv g(n\Delta t)$ to write
\begin{equation}
  \hat{g}_{obs}(f) = \sum_{n=0}^N g_n e^{-2\pi i f n \Delta t}
  \eqlabel{DFT-f}
\end{equation}
From the arguments around Nyquist aliasing, we know that the only relevant
frequency range is from $0 \le f \le 1/\Delta t$, and so we can define $N$
evenly-spaced frequencies with $\Delta f = 1 / (N\Delta t)$ covering this range.
Denoting the sampled transform as
$\hat{g}_k \equiv \hat{g}_{obs}(k\Delta f)$, we can write
\begin{equation}
  \hat{g}_k = \sum_{n=0}^N g_n e^{-2\pi i k n / N}
  \eqlabel{DFT}
\end{equation}
which you might recognize as the standard form of the discrete Fourier
transform.

Notice, though, that we glossed over one important thing: the effect of switching
from an infinite number of samples to a finite number of samples.
In moving from \eq{DFT-f-inf} to \eq{DFT-f}, we have effectively applied
to our data a rectangular window function of width $N\Delta t$.
From the discussion accompanying \fig{rectangular-window}, we know what this
does: it gives us a Fourier transform convolved with a sinc function of width
$1 / (N\Delta t)$, resulting in the ``smearing'' of the Fourier transform signal
with this width.
Roughly speaking, then, any two Fourier transform values at frequencies within
$1/(N\Delta t)$ of each other will not be independent, and so we should space
our evaluations of the frequency with $\Delta f \ge 1/(N\Delta t)$.
Comparing to above, we see that this is {\it exactly the frequency spacing}
we arrived at from Nyquist-frequency arguments.

What this indicates is that the frequency spacing of the discrete Fourier
transform is optimal in terms of both the Nyquist sampling limit
{\it and} the effect of the finite observing window!
Now, this argument has admittedly been a bit hand-wavy, but there do exist
mathematically rigorous approaches to proving that the discrete Fourier
transform in \eq{DFT} captures all of the available frequency information
for a uniformly-sampled function $g_n$
\citep[see, e.g.][]{FoundationsOfSignalProcessing}.
Despite our lack of rigor here, I find this to be a helpful approach
in developing intuition regarding the relationship between the
continuous and discrete Fourier transforms.

\subsection{The Classical Periodogram}

With the discrete Fourier transform defined in \eqs{DFT-f}{DFT}, we can
apply the definition of the Fourier power spectrum from \eq{power-spectrum}
to compute the {\it classical periodogram}, sometimes called the
{\it Schuster periodogram} after \citet{Schuster98} who first proposed it:
\begin{equation}
  P_S(f) = \frac{1}{N}\left|\sum_{n=1}^N g_n e^{-2\pi i f t_n}\right|^2
  \eqlabel{schuster-periodogram}
\end{equation}
Apart from the $1/N$ proportionality, this sum is precisely the Fourier power
spectrum in \eq{power-spectrum}, computed for a continuous signal observed
with uniform sampling defined by a Dirac comb.
It follows that, in the uniform sampling case, the Schuster periodogram
captures all of the relevant frequency information present in the data.
This definition readily generalizes to the non-uniform case, which we
will explore in the following section.

One point that should be emphasized is that the {\it periodogram} in
\eq{schuster-periodogram} and the {\it power spectrum} in \eq{power-spectrum}
are conceptually different things.
As noted in \citet{Scargle82}, the astronomy community tends to use these
terms interchangeably, but to be precise
the periodogram---i.e., the statistic we compute from our data---is an
{\it estimator} of the power spectrum---i.e., the underlying continuous
function of interest.
In fact, the classical periodogram and its extensions (including the
Lomb-Scargle we will discuss momentarily)
are not consistent estimators of the power spectrum: that is, the periodogram
has unavoidable intrinsic variance, even in the limit of an infinite number
of observations
\citep[for a detailed discussion, see Chp 8.4 of][]{Anderson1971}.

\section{Non-uniform Sampling}
\sectlabel{non-uniform-sampling}

In the real world, particularly in fields like Astronomy where observations are
subject to influences of weather and diurnal, lunar, or seasonal cycles, the
sampling rate is generally far from uniform.
Using the same approach as we used to explore uniform sampling in the previous
section, we can now explore non-uniform sampling here.

In the general non-uniform case, we measure some signal at a set of $N$ times
which we will denote $\{t_n\}$, which leads to the following observing window:
\begin{equation}
  W_{\{t_n\}}(t) = \sum_{n=1}^{N} \delta(t - t_n)
  \eqlabel{nonuniform-window}
\end{equation}
Applying this window to our true underlying signal $g(t)$, we find an observed
signal of the form:
\begin{eqnarray}
  g_{obs}(t) &=& g(t) W_{\{t_n\}}(t) \nonumber\\
             &=& \sum_{n=1}^{N} g(t_n)\delta(t - t_n)
  \eqlabel{g-nonuniform}
\end{eqnarray}

Just as in the evenly-sampled case, the Fourier transform of the observed
signal is a convolution of the transforms of the true signal and the window:
\begin{equation}
  \mathcal{F}\{g_{obs}\} = \mathcal{F}\{g\} \ast \mathcal{F}\{W_{\{t_n\}}\}
  \eqlabel{g-obs-conv}
\end{equation}
Unlike in the uniform case, the window transform $\mathcal{F}\{W_{\{t_n\}}\}$
will generally {\it not} be a straightforward sequence of delta functions; the
symmetry present in the Dirac comb is broken by the uneven sampling,
leading the transform to be much more ``noisy''.
This can be seen in \fig{random-window}, which shows the Fourier transform of
a non-uniform observing window with an average sampling rate identical to
that in \fig{comb-window-1}, along with its imact on the observed Fourier
transform.

\begin{figure}[ht]
  \centering
  \includegraphics[width=\textwidth]{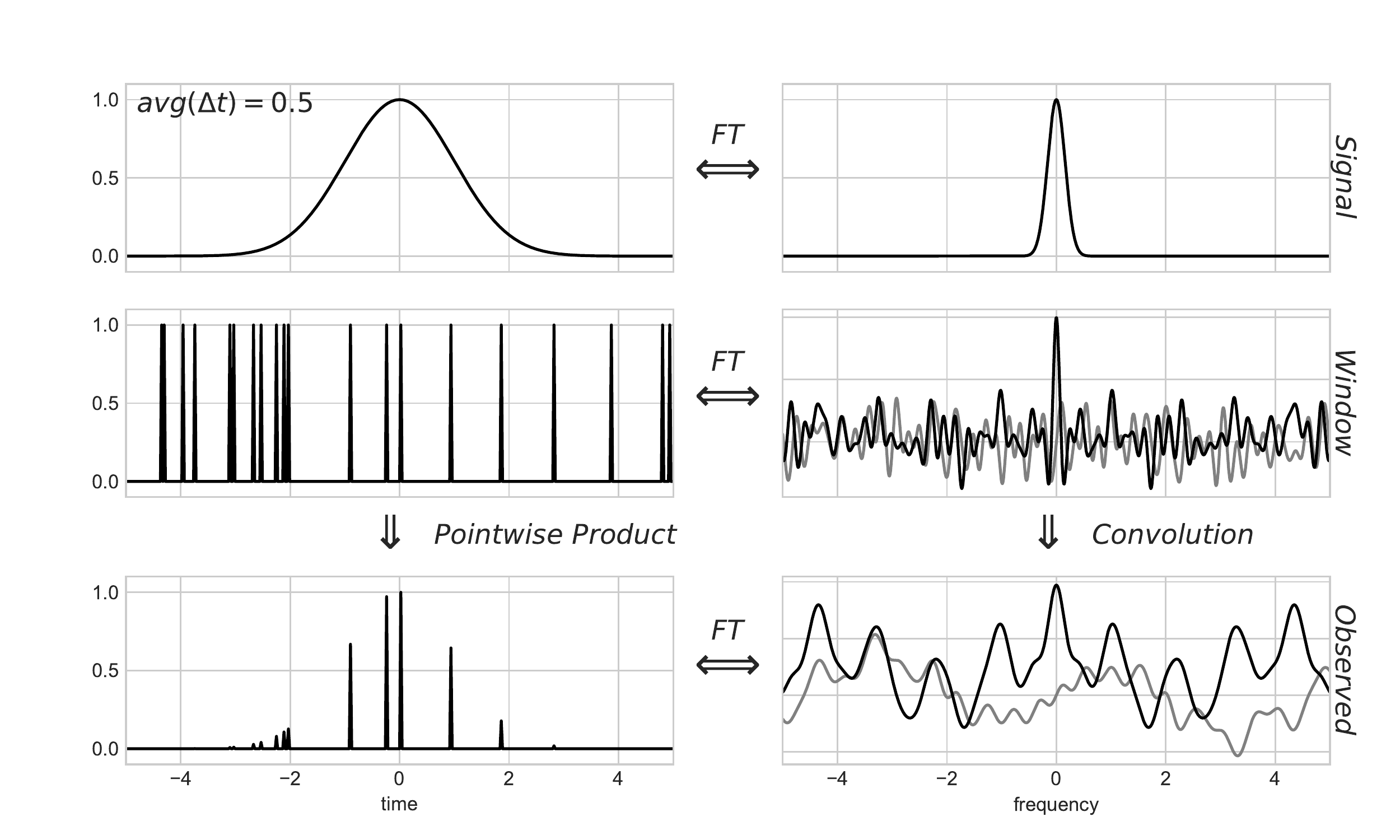}
  \caption{The effect of non-uniform sampling on the observed Fourier transform.
    These samples have the same average spacing as those in \fig{comb-window-1},
    but the irregular spacing within the observing window translates to
    irregular frequency peaks in its transform,
    causing the observed transform to be ``noisy''.
    Here black and gray lines represent the real and imaginary parts of the
    transform, respectively.
    \figlabel{random-window}}
\end{figure}

A few things stand-out in this figure. In particular, the Fourier transform of
the non-uniformly spaced delta functions looks like random noise, and in some
sense it is: the locations and heights of the Fourier peaks are related to
the intervals between observations, and so randomization of observation times
leads to a randomization of Fourier peak locations and heights.
That is, {\it non-structured spacing of observations will lead to
a non-structured frequency peaks in the window transform}.
This non-structured window transform, when convolved with the Fourier transform
of the true signal, results in an observed Fourier transform reflecting the
same random noise.
Comparing to the uniformly-spaced observations in \fig{comb-window-1}, we
see that the unstructured nature of the window transform means that there
is no exact aliasing of the true signal, and thus no way to exactly recover
any portion of the true Fourier transform for the underlying function.

\begin{figure}[ht]
  \centering
  \includegraphics[width=\textwidth]{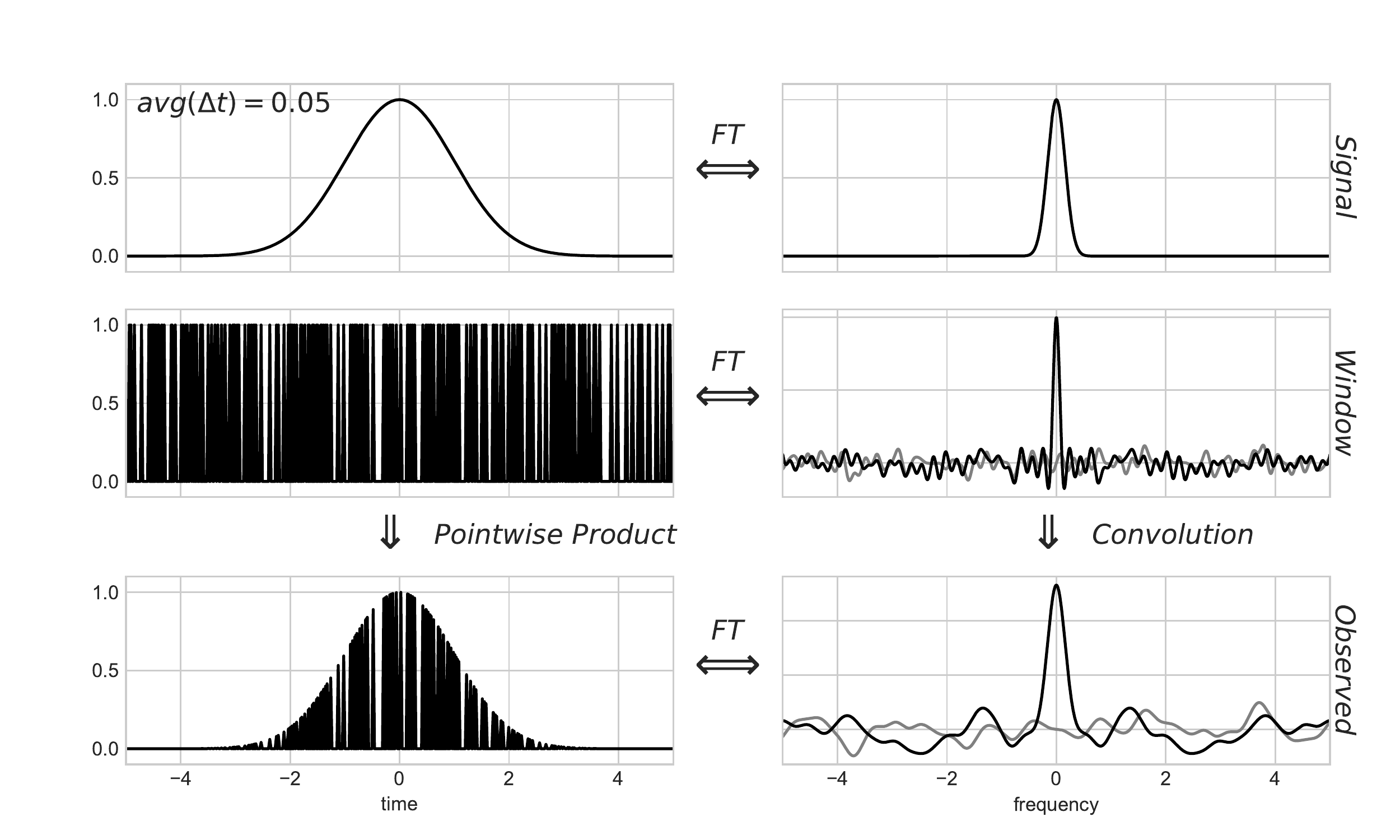}
  \caption{The effect of non-uniform sampling on the observed Fourier transform,
    with a factor of 10 more samples than \fig{random-window}.
    Even with very dense sampling of the function, the Fourier transform
    cannot be exactly recovered due to the imperfect aliasing present in
    the window transform.
    \figlabel{random-window-2}}
\end{figure}

One might hope that sampling the signal more densely might alleviate
these problems, and it does, but only to a degree.
In \Fig{random-window-2} we increase the density of observations by a factor of 10,
such that there are 200 total observations over the length-10 observing window.
The observed Fourier transform in this case is much more reflective of the
underlying signal, but still contains a degree of ``noise'' rooted in the
randomized frequency peaks due to randomized spacing between observations.

\subsection{A Non-uniform Nyquist Limit?}
\sectlabel{pseudo-nyquist}

We saw in \sect{nyquist} that the Nyquist limit is a direct consequence
of the symmetry in the Dirac comb window function that describes evenly-sampled data,
and uneven sampling destroys the symmetry that underlies its definition.
Nevertheless, the idea of the ``Nyquist frequency'' seems to have taken hold
in the scientific psyche to the extent that the idea is often mis-applied in
areas where it is mathematically irrelevant.
For unevenly-sampled data, the truth is that the ``Nyquist limit'' might or
might not exist, and even in cases where it does exist it tends to be far
larger (and thus far less relevant) than in the evenly-sampled case.

\subsubsection{Incorrect Limits in the Literature}

In the scientific literature it is quite common to come across various proposals
for a Nyquist-like limit applied in the case of irregular sampling.
A few typical approaches include using the mean of the sampling intervals
\citep[e.g.][]{Scargle82, Horne86, NumRec},
the harmonic mean of the sampling intervals \citep[e.g.][]{Debosscher07},
the median of the sampling intervals \citep[e.g.][]{Graham2013b},
or the minimum sample spacing \citep[e.g.][]{Percy86, Roberts87, Press89, Hilditch01}.
All of these ``pseudo-Nyquist'' limits are tempting criteria in that
they are easy to compute, and reduce to the classical Nyquist
frequency in the limit of evenly-spaced data.
Unfortunately, none of these approaches is correct: in general,
unevenly-sampled data can probe frequencies far larger than any of these
supposed limits (a fact that several of these citations do hint at
parenthetically).

As a simple example of where such logic can fail spectacularly,
consider the data from \fig{LINEAR-data}: though the mean sample
spacing is one observation every seven days, in \fig{LINEAR-power}
we were nevertheless able to quite clearly identify a period of 2.58 hours---
an order of magnitude shorter than the average-based pseudo-Nyquist limit
would indicate as possible.
For the data in \fig{LINEAR-data}, the minimum sample spacing is just under 10
seconds, but it would be irresponsible to claim that this single pair of
observations by itself defines some limit beyond which frequency information
is unattainable.

\begin{figure}[ht]
  \centering
  \includegraphics[width=\textwidth]{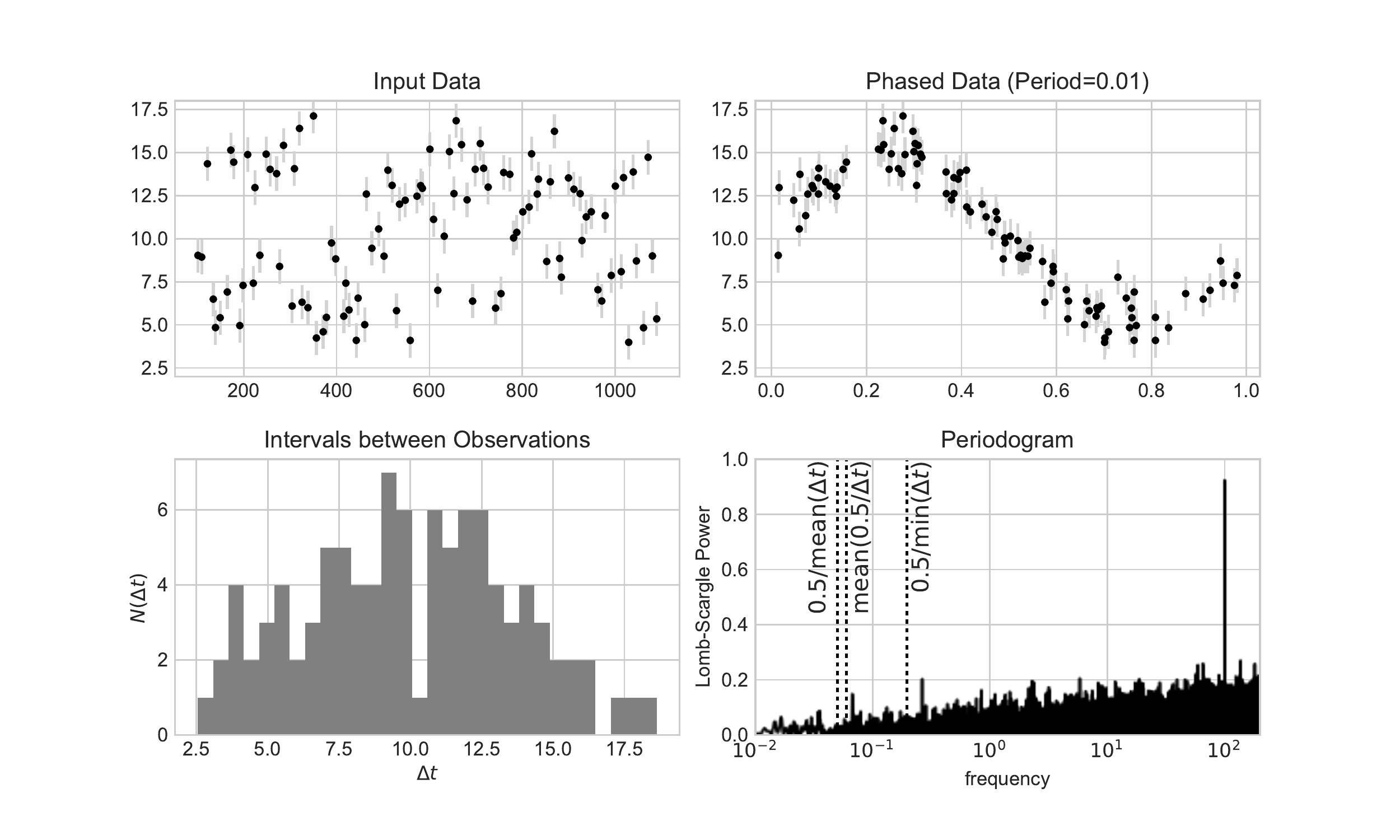}
  \caption{An example of data for which the various poorly-motivated
    ``pseudo-Nyquist'' approaches outlined in \sect{pseudo-nyquist} fail
    spectacularly. The upper panels show the data, a noisy sinusoid with
    a frequency of 100 (i.e. a period of 0.01).
    The lower left panel shows a histogram of spacings between observations:
    the minimum spacing is 2.55, meaning that the signal has
    {\it over 250 full cycles} between the closest pair of observations.
    Nevertheless, the periodogram (lower right) clearly identifies the correct
    period, though it is orders of magnitude larger than pseudo-Nyquist
    estimates based on average or minimum sampling rate.
    \figlabel{pseudo-nyquist}}
\end{figure}

As a more extreme example, consider the data shown in \fig{pseudo-nyquist}.
This consists of noisy samples from a sinusoid with a period of 0.01 units,
with sample spacings ranging between 2 and 18 units: needless to say, any
pseudo-Nyquist definition based on an average or minimum sample spacings
will be {\it far} below the true frequency of 100; still, the
the Lomb-Scargle periodogram in the lower right panel quite cleanly
recovers the true frequency.

\subsubsection{The Non-uniform Nyquist Limit}
\sectlabel{non-uniform-nyquist}

While pseudo-Nyquist arguments based on average or minimum sampling
fail spectacularly, there is a sense in which the Nyquist limit can
be applied to unevenly-spaced data. \citet{Eyer99} explore this issue
in some detail, and in particular prove the following:
\begin{quote}
{\it Let $p$ be the largest value such that each $t_i$ can be written
  $t_i = t_0 + n_i p$, for integers $n_i$. The Nyquist Frequency then
  is $f_{Ny} = 1 / (2p)$.}
\end{quote}
In other words, computing the Nyquist limit for unevenly-spaced data requires
finding the largest factor $p$, such that each spacing $\Delta t_i$ is
{\it exactly} an integer multiple of this factor.
\citet{Eyer99} prove this formally, but the result can be understood
by thinking of such data as a windowed version of uniformly sampled
data with spacing $p$, where the window is zero at all points but the
location of the observations.
Such uniform data has a classical Nyquist limit of $1/(2p)$, and a window
function applied on top of that sampling does not change that fact.

\Fig{nyquist-eyer99} shows an example of such a Nyquist frequency.
The data are non-uniformly sampled at times $t_i = n_i \cdot p$, with $p=0.01$
and $n_i$ drawn randomly from positive integers less than 10,000.
According to the \citet{Eyer99} definition,
this results in a Nyquist frequency $f_{Ny} = 50$, and we see the expected
behavior beyond this frequency: the signal at $f > f_{Ny}$ consists of a series
of exact aliases of the signal at $|f| < f_{Ny}$.

\begin{figure}[ht]
  \centering
  \includegraphics[width=\textwidth]{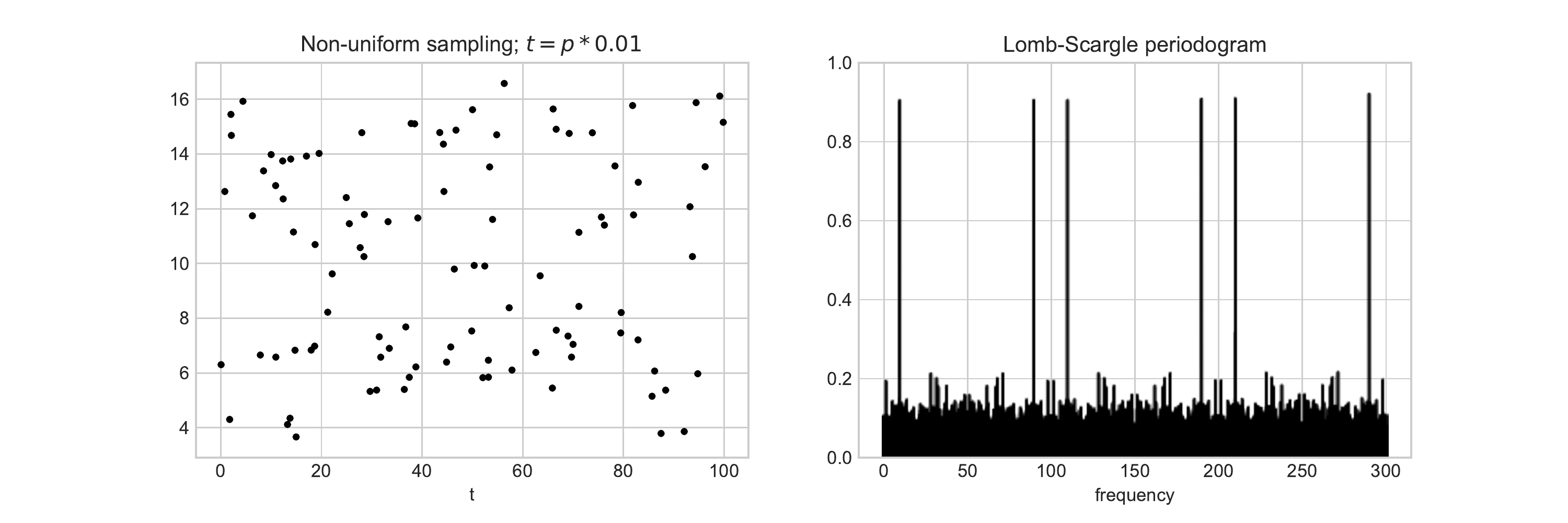}
  \caption{A visualization of the \citet{Eyer99} definition of the Nyquist
    frequency. Data are non-uniformly sampled at times $t_i = n_i p$, for
    integer $n_i$ and $p=0.01$.
    This results in a Nyquist frequency of $f_{Ny}= (2p)^{-1} = 50$:
    the periodogram outside the range $0 \le f < f_{Ny}$ is a series of
    perfect aliases of the signal within that range.
    \figlabel{nyquist-eyer99}}
\end{figure}

We should keep in mind one consequence of this Nyquist definition:
if you have any pair of observation spacings
whose ratio is irrational, {\it the Nyquist limit does not exist!}.
To realize this situation in practice, however,
would require infinitely precise measurements of the
times $t_i$; finite precision of time measurements means the Nyquist
frequency can be large, but not infinite.
For example, if your observation times are recorded to $D$ decimal places,
the Nyquist frequency will be at most
\begin{equation}
  f_{Ny} \le \frac{1}{2} 10^D,
  \eqlabel{nonuniform-nyquist}
\end{equation}
with the inequality due to the fact that larger common factors may exist.
In other words, absent other relevant patterns in the observations,
the Nyquist frequency for irregularly-sampled data is most
typically set by {\it the precision of the time measurements}
\citep[see also][for more rigorous treatments
  of this result]{Bretthorst2003, Koen2006}.

\subsubsection{Frequency Limit due to Windowing}

In contexts where observations are not instantaneous, but rather consist of
short-duration integrations of a continuous signal, a qualitatively different
kind of frequency limit exists.
This is typical in, e.g., optical astronomy, where a single observation typically consists of an integration of observed photons over a finite duration $\delta t$.
As noted by \citet{ICVG2014}, this time-scale of integration represents another
kind of limiting frequency for irregularly-sampled data.
Such a situation means that the observation is effectively a convolution of
the underlying signal with a rectangular window function of width $\delta t$,
in a manner analogous to \fig{convolution-theorem}.
By the convolution theorem, the observed transform will be a point-wise
product between the true transform and the transform of the window, which
will generally have a width proportional to $1/\delta t$.
This means that---absent other more constraining window effects---the
frequency limit is $f_{max} \propto 1/(2\delta t)$, with the constant of
proportionality dependent on the shape of the effective window describing
individual observations.

Keep in mind that the windowing limit $1/(2\delta t)$ is quite different than a
Nyquist limit: the Nyquist limit is the frequency beyond which all signal
is aliased into the Nyquist range; the windowing limit is the frequency
beyond which all signal is attenuated to zero.
In practice, the limit implied by either the temporal resolution or windowing
of individual observations may be too large to be computationally feasible;
for discussion of frequency limits in practice, see \sect{frequency-grid}.

\subsection{Semi-structured Observing Windows}

We have seen that for uniform data, the perfect aliasing beyond the Nyquist
frequency is a direct consequence of the symmetry of the Dirac-comb window
function.
For non-uniform observations, such symmetry does not exist, but {\it structure}
in the observing window can lead to partial aliasing of signals in the
data \citep[see, e.g.][]{Deeming75}.
In this section, we will examine two typical window functions derived from
real-world observations: one ground-based (LINEAR) and one space-based (Kepler).
For details on how window power spectra can be estimated in practice, see
\sect{windows-and-deconvolution}.

\begin{figure}[ht]
  \centering
  \includegraphics[width=\textwidth]{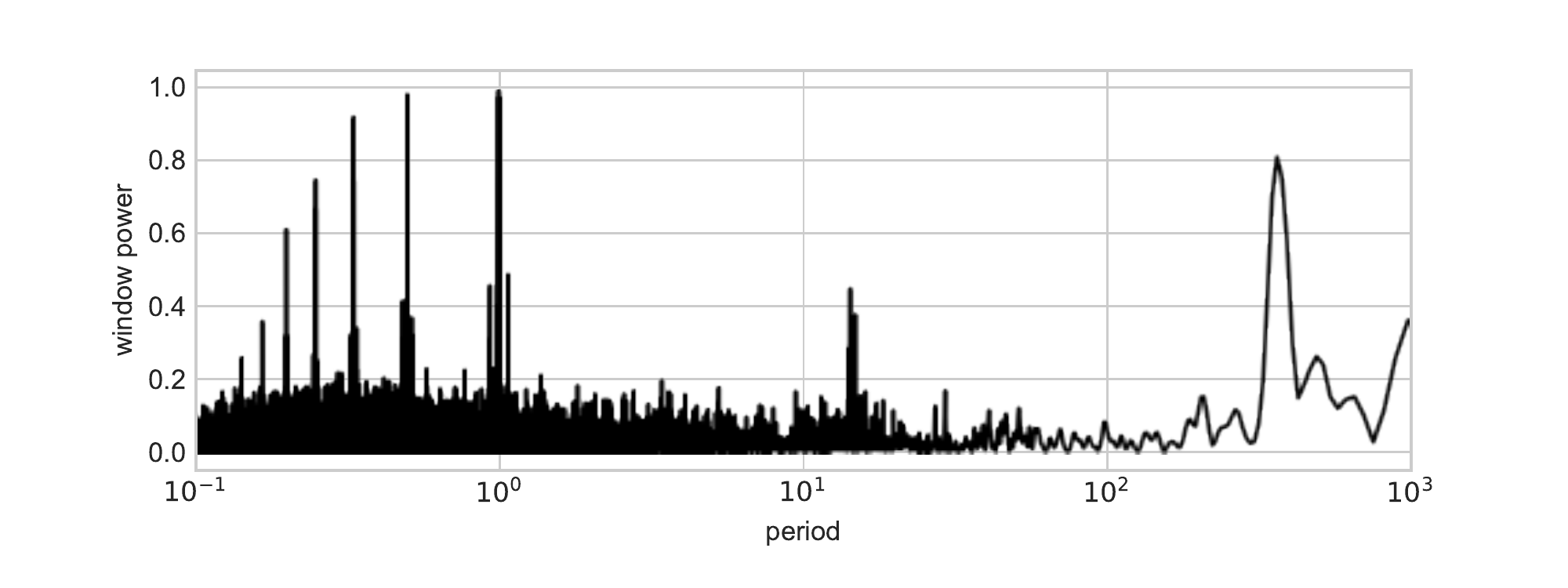}
  \caption{The power spectrum of the observing window for the data shown
    in \fig{LINEAR-data}. Notice the strong spike in power at a period of
    1 day, and related aliases at $1/n$ days for integer $n$.
    There is also a strong spike at $365$ days, and a noticeable spike at
    $\sim 14$ days. Each of these indicate time intervals that appear often
    in the data.
    \figlabel{LINEAR-window}}
\end{figure}

\subsubsection{A Ground-based Observing Window: LINEAR}

Let's again consider the data shown in \fig{LINEAR-data}. The window power
spectrum for this in \fig{LINEAR-window} shows some quite distinct features,
and these features have an intuitive interpretation.
Namely, if the window power shows a spike at a period of $p$ days, this means
that an observation at time $t_0$ is likely to be followed by another
observation near a time $t_0 + np$ for integer $n$.

With this in mind, the strong spike at a period of 1 day indicates
that observations are
taken near the same time of day: this is typical of a ground-based survey
with observations recorded only during the nighttime hours.
The additional spikes at periods of $1/n$ days (for integer $n$) are aliases
of this same feature.
\fig{LINEAR-window} also shows a wide spike at a period of 1 year, indicating
a detectable annual pattern in the observations.
Finally, there is a noticeable spike at approximately 14 days that is likely
related to patterns of scheduling within the survey.

Recall that a Fourier spectrum observed through a particular window will reflect
a convolution of the true spectrum and the window spectrum
(cf.\ \fig{random-window}), and so we would expect the structure in the window
to be imprinted on the power spectrum measured from the data.

\begin{figure}[ht]
  \centering
  \includegraphics[width=\textwidth]{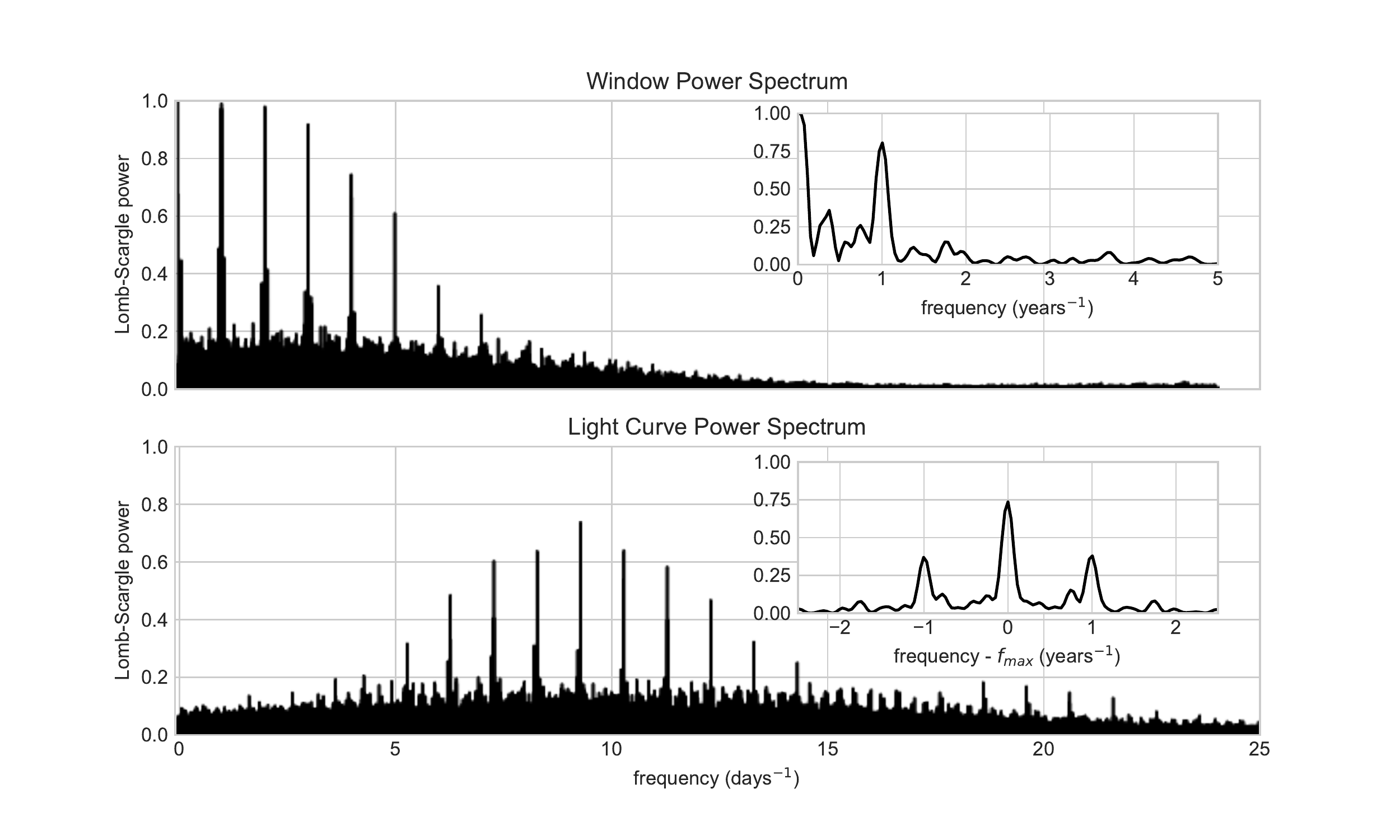}
  \caption{The effect of the window function in \fig{LINEAR-window} on the
    power spectrum in \fig{LINEAR-power}.
    The top panel shows the window power spectrum, and the bottom panel shows
    the observed signal power spectrum.
    Both are plotted as a function of frequency (we earlier saw both of these
    as a function of period; see \fig{LINEAR-window} and \fig{LINEAR-power},
    respectively).
    Viewing these as a function of frequency makes it clear that the structure
    in the window function is imprinted on the observed spectrum: both the
    diurnal structure in the main panel, and the annual structure in the inset
    are apparent in the observed spectrum.
    \figlabel{LINEAR-window-effect}}
\end{figure}

This imprint of the window power is illustrated in \fig{LINEAR-window-effect}.
The upper panel shows the window power spectrum as a function of frequency
(rather than period, as in \fig{LINEAR-window}),
while lower panel shows the observed
signal power spectrum as a function of frequency
(rather than period, as in \fig{LINEAR-power}).
The upper panel and its inset show clearly the 1-day and 1-year features we
noted previously.
The lower panel shows the observed power spectrum of the data: these diurnal
and annual peaks in the window function are quite clearly imprinted on the
observed power spectrum at relevant scales.
This approximate aliasing is similar to the exact aliasing seen in
regularly-sampled data at the Nyquist frequency; however, in this case
the magnitude of the aliased signal fades further from the frequency
driving the signal.

\subsubsection{A Space-based Observing Window: Kepler}

\begin{figure}[ht]
  \centering
  \includegraphics[width=\textwidth]{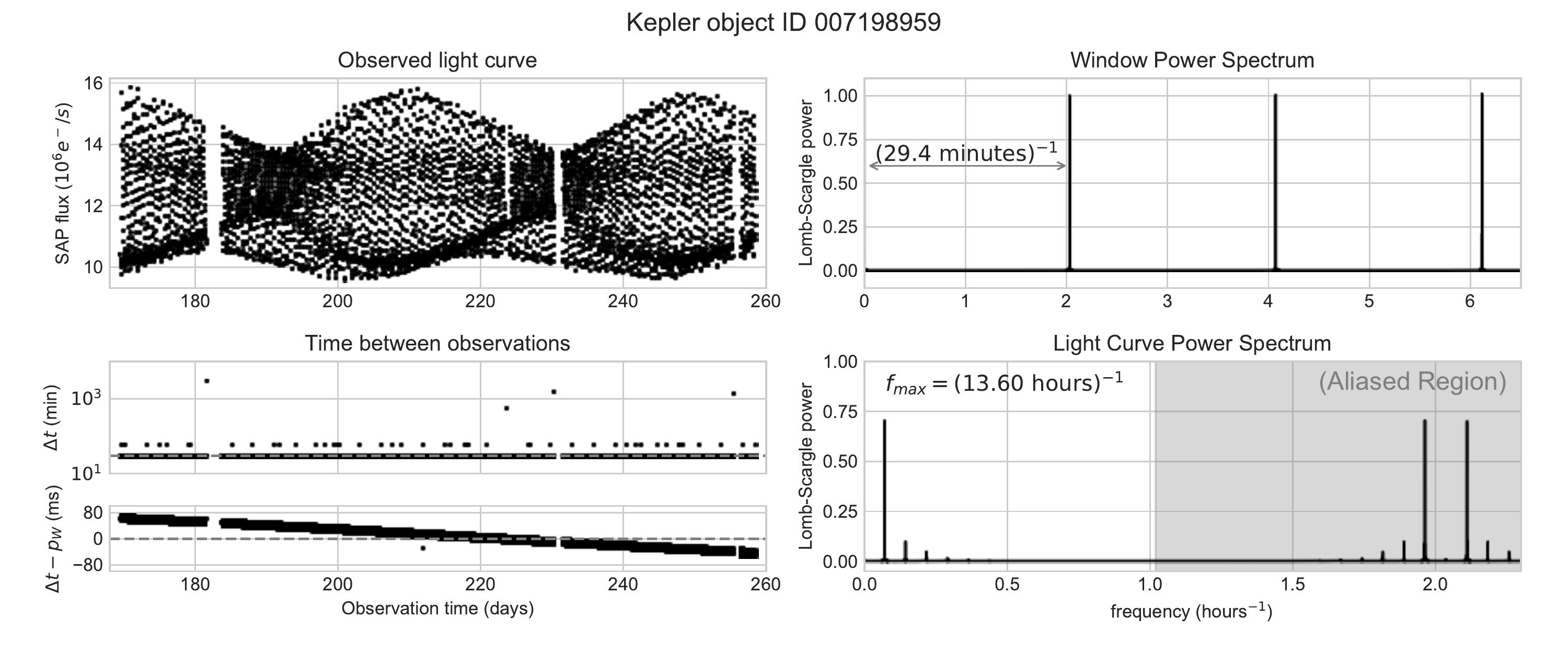}
  \caption{An RR Lyrae variable observed by the Kepler project \citep[see][]{Kolenberg2010}.
    {\it upper left:} the 4083 observed fluxes over roughly three months.
    {\it upper right:} the window power spectrum, which is quite close to that of regularly-spaced data with a cadence of 29.4 minutes.
    {\it lower left:} the time between observations. Missing measurements aside, the spacing between observations is nearly uniform. the lower panel gives a closer look at the majority of the spacings, which are not exactly the same but rather span a range of $\pm 50$ms around the 29.4-minute period observed in the window function.
    {\it lower right:} the data power spectrum, showing approximate aliasing and clearly indicating a peak near a period of 13.6 hours, along with higher-order components at multiples of this value.
    \figlabel{kepler-data}}
\end{figure}

Space-based surveys will generally have quite different observing windows.
For example, the upper-left panel of \fig{kepler-data}
shows observations of an RR-Lyrae variable star from the Kepler survey,
measured 4083 times over a period of three months,
with an irregular observing cadence of around 30 minutes
 \citep[For deeper discussion of these observations, see][]{Kolenberg2010}.
The Kepler observations are very nearly uniformly-spaced, and this is reflected
in the window power spectrum, shown in the upper-right panel of
\fig{kepler-data}.
The window function is a series of very narrow evenly-spaced spikes,
reminiscent of the Dirac comb shown in \figs{comb-window-1}{comb-window-2}.
By analogy we can treat $f_{Ny} = 0.5/29.4$ minutes$^{-1}$ as the
effective Nyquist limit for the data, keeping in mind that aliasing beyond this
``limit'' will be imperfect due to the uneven spacing of the samples
(see the lower-left panels of \fig{kepler-data}).
The lower-right panel of \fig{kepler-data} shows the power spectrum of the
observations, with gray shading indicating the (nearly) aliased region
of the spectrum.
The period of 13.6 hours is quite strongly apparent, along with smaller spikes
at integer multiples of this frequency that indicate higher-order periodic
components in the signal.

The window functions for ground-based and space-based observations, reflected
by LINEAR data in \fig{LINEAR-window-effect} and Kepler data in
\fig{kepler-data}, are quite different, but in both cases essential features
of the observed power spectra can be understood by recognizing that the
periodogram measures
not the power spectrum of the underlying signal, but a power specrtrum from
the convolution of the true signal transform and the Fourier transform of
the window function.

\section{From Classical to Lomb-Scargle Periodograms}
\sectlabel{schuster-to-lomb-scargle}

Up until now, we have been mainly discussing the direct extension of the
classical periodogram in \eq{schuster-periodogram} to non-uniform data.
Returning to this definition, we can rewrite the expression in a more
suggestive way:
\begin{eqnarray}
  P(f)
  &=& \frac{1}{N}\left|\sum_{n=1}^N g_n e^{-2\pi i f t_n} \right|^2 \nonumber\\
  &=& \frac{1}{N}\left[
    \left(\sum_n g_n \cos(2\pi f t_n)\right)^2
    + \left(\sum_n g_n \sin(2\pi f t_n)\right)^2
    \right]
  \eqlabel{classical-periodogram}
\end{eqnarray}
Although this form of the non-uniform periodogram can be useful for identifying
periodic signals, its statistical properties are not as straightforward as in
the uniform case.
When the classical periodogram is applied to uniformly-sampled Gaussian noise,
the values of the resulting periodogram is chi-square distributed.
This property becomes quite useful in practice when the periodogram is used in the
context of a classical hypothesis test to distinguish between periodic and
non-periodic objects---see \sect{false-alarm-probability}.
Unfortunately, when the sampling becomes nonuniform these properties no longer
hold and the periodogram distribution cannot in general be analytically expressed.

\citet{Scargle82} addressed this by considering a generalized form of the periodogram,
\begin{equation}
  P(f) = \frac{A^2}{2}\left(\sum_n g_n \cos(2\pi f [t_n-\tau])\right)^2
       + \frac{B^2}{2} \left(\sum_n g_n \sin(2\pi f [t_n-\tau])\right)^2,
\end{equation}
where $A$, $B$, and $\tau$ are arbitrary functions of the frequency $f$ and
observing times $\{t_i\}$ (but not the values $\{g_n\}$), and showed
that you can choose a unique form of $A$, $B$, and $\tau$ such that
\begin{enumerate}
  \item The periodogram reduces to the classical form in the case of equally-spaced observations,
  \item The periodogram's statistics are analytically computable,
  \item The periodogram is insensitive to global time-shifts in the data.
\end{enumerate}
The values of $A$ and $B$ leading to these properties result in the following
form of the generalized periodogram:
\begin{eqnarray}
  P_{LS}(f) =
  \frac{1}{2} \Bigg\{ &
  \bigg(\sum_n g_n \cos(2\pi f [t_n-\tau])\bigg)^2 \bigg/
  \sum_n \cos^2(2\pi f [t_n-\tau]) &\nonumber\\
  & + ~ \bigg(\sum_n g_n \sin(2\pi f [t_n-\tau])\bigg)^2 \bigg/
  \sum_n \sin^2(2\pi f [t_n-\tau]) & \Bigg\}
  \eqlabel{lomb-scargle-periodogram}
\end{eqnarray}
where $\tau$ is specified for each $f$ to ensure time-shift invariance:
\begin{equation}
  \tau = \frac{1}{4\pi f}\tan^{-1}\Bigg(
  \frac{\sum_n \sin(4\pi f t_n)}{\sum_n \cos(4\pi f t_n)}\Bigg).
  \eqlabel{tau-def}
\end{equation}
This modified periodogram differs from the classical periodogram only to
the extent that the denominators $\sum_n \sin^2(2\pi f t_n)$ and
$\sum_n \cos^2(2\pi f t_n)$ differ from $N/2$, which is the expected value of
each of these quantities in the limit of complete phase sampling at each
frequency.
Thus, in many cases of interest the Lomb-Scargle periodogram only differs
slightly from the classical/Schuster periodogram; an example of this is seen
in \fig{ls-comparison}.

\begin{figure}[ht]
  \centering
  \includegraphics[width=\textwidth]{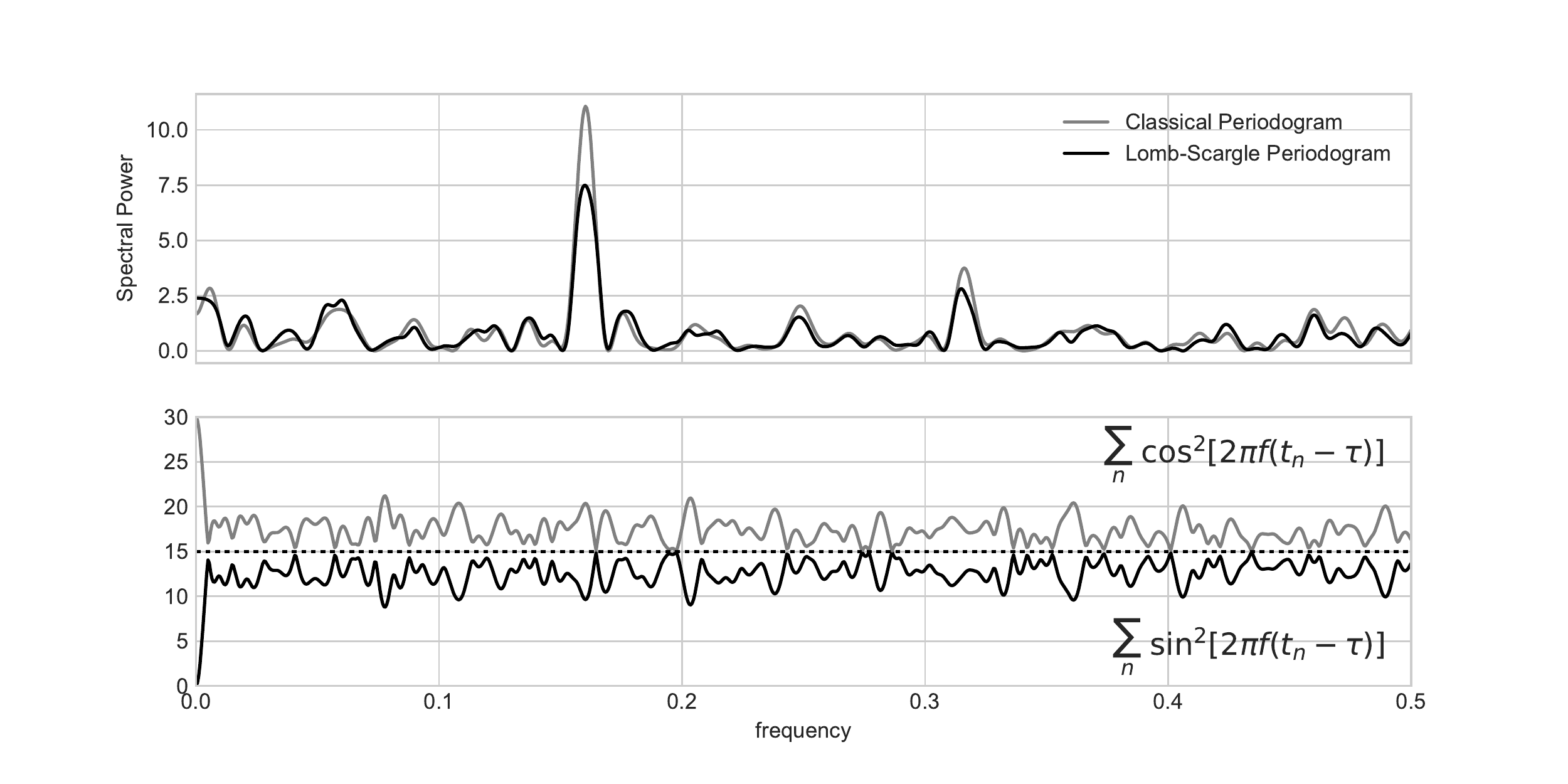}
  \caption{{\it upper panel:} A comparison of the Classical periodogram
    (\eq{classical-periodogram}) and the Lomb-Scargle periodogram
    (\eq{lomb-scargle-periodogram}) for 30 noisy points drawn from a sinusoid.
    Though the two periodogram estimates differ quantitatively, the essential
    qualitative features---namely the position of significant peaks---typically
    remain the same.
    {\it lower panel:} the values of the denominators in
    \eq{lomb-scargle-periodogram}.
    The difference between the Lomb-Scargle periodogram and the classical
    periodogram stems from the difference between these quantities
    and $N/2 = 15$ (the dotted line).
    \figlabel{ls-comparison}}
\end{figure}

A remarkable feature of Scargle's modified periodogram is that it is
{\it identical} to the result
obtained by fitting a model consisting of a simple sinusoid to the data at
each frequency $f$ and
constructing a ``periodogram'' from the $\chi^2$ goodness-of-fit at each
frequency---an estimator which was considered in some depth by \citet{Lomb76}.
From this perspective, the $\tau$ shift defined in \eq{tau-def} serves to
orthogonalize the normal equations used in the least squares analysis.
Partly due to this deep connection between Fourier analysis and least-squares
analysis, the modified periodogram in \eq{lomb-scargle-periodogram}
has since become commonly referred to as the {\it Lomb-Scargle Periodogram},
although versions of this approach had been employed even earlier
\citep[see, e.g.][]{Gottlieb75}.

Because of the close similarity between the classical and Lomb-Scargle periodograms,
the bulk of or previous discussion applies---at least qualitatively---to
periodograms computed with the Lomb-Scargle method.
In particular, reasoning about the effect of window functions on the observed
Lomb-Scargle power spectrum remains qualitatively useful even if it is not
quantitatively precise.

One important caveat of the simple Lomb-Scargle formula is that the statistical
guarantees only apply when the observations have {\it uncorrelated white
noise}; data with more complicated noise characteristics must be treated
more carefully; see, e.g., \citet{Vio2010} or the Least Squares approach
discussed in \sect{extensions-observational-noise}.

\section{The Least-Square Periodogram and its Extensions}
\sectlabel{lomb-scargle-extensions}

The equivalence of the Fourier interpretation and least squares interpretation of the Lomb-Scargle periodogram allows for some interesting and useful
extensions, some of which we will explore in this section.
First, let's consider the Least Squares periodogram itself.

In the least squares interpretation of the periodogram, a sinusoidal model is proposed at each candidate frequency $f$:
\begin{equation}
  y(t;f) = A_f \sin(2 \pi f (t - \phi_f))
\end{equation}
where the amplitude $A_f$ and phase $\phi_f$ can vary as a function of frequency.
These model parameters are fit to the data in the standard least-squares sense,
by constructing the $\chi^2$ statistic at each frequency:
\begin{equation}
  \chi^2(f) \equiv \sum_n \big(y_n - y(t_n;f)\big)^2
  \eqlabel{chi2-simple}
\end{equation}
We can find the ``best'' model $\hat{y}(t;f)$ by minimizing $\chi^2(f)$ at
each freuqnecy with respect to $A_f$ and $\phi_f$;
we will denote this minimum value as $\hat{\chi}^2(f)$.
\citet{Scargle82} showed that with this setup, the Lomb-Scargle
periodogram from \eq{lomb-scargle-periodogram} can be equivalently written:
\begin{equation}
  P(f) = \frac{1}{2}\big[\hat{\chi}^2_0 - \hat{\chi}^2(f)\big]
  \eqlabel{lomb-scargle-chi2}
\end{equation}
where $\hat{\chi}^2_0$ is the non-varying reference model.
The key realization here is that the Lomb-Scargle periodogram essentially
{\it assumes a sinusoidal model} for the data; this is visualized in
\fig{ls-model} for the data we had seen in \fig{LINEAR-data}.
This immediately begs the question: can we compute a ``periodogram'' based
on more general forms of the above model to more effectively fit the data?

\begin{figure}[ht]
  \centering
  \includegraphics[width=\textwidth]{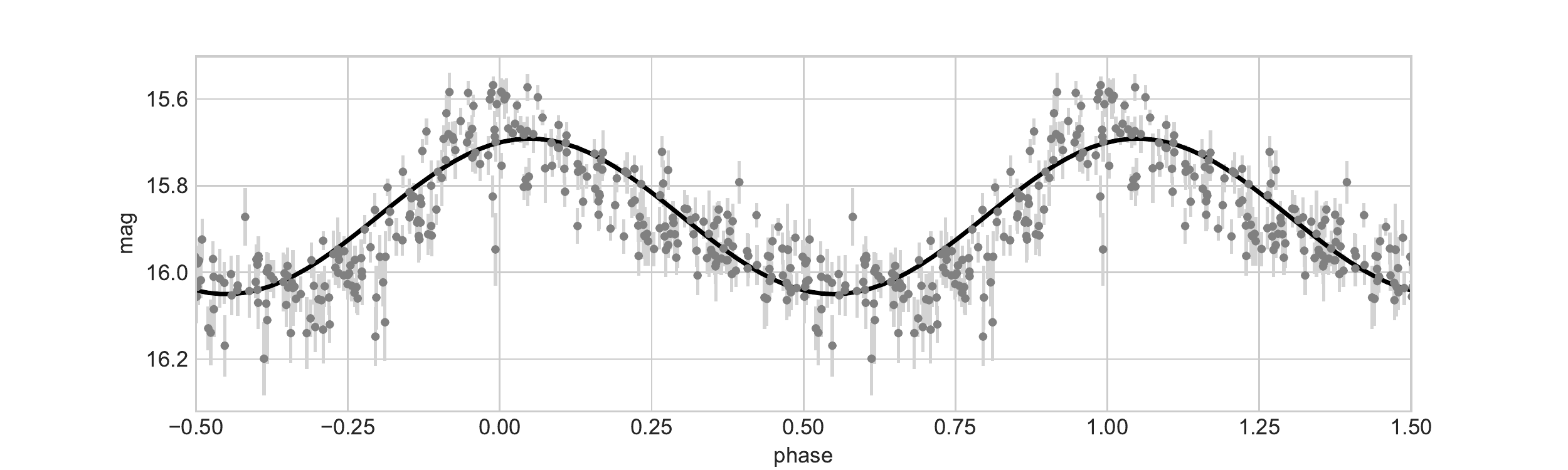}
  \caption{The sinusoidal model implied by the Lomb-Scargle periodogram for
    the LINEAR data seen in \fig{LINEAR-data}.
    Although a sinusoid does not perfectly fit the data, the sinusoidal model
    is close enough that it serves to locate the correct frequency.
    \figlabel{ls-model}}
\end{figure}

\subsection{Measurement Errors}
\sectlabel{extensions-observational-noise}
Perhaps the most important modification to the periodogram is to
construct it such that it correctly handles measurement error in the data.
This can be done through the standard change to the $\chi^2$ expression:
i.e., if there are Gaussian errors $\sigma_n$ on each observed $y_n$,
we can re-write \eq{chi2-simple} in the following (standard) way:
\begin{equation}
  \chi^2(f) \equiv \sum_n \left(\frac{y_n - y_{model}(t_n;f)}{\sigma_n}\right)^2
  \eqlabel{chi2-with-errors}
\end{equation}
The periodogram constructed from this $\chi^2$ definition will more accurately
reflect the spectral power of noisy observations.
The effect of this modification on \eq{lomb-scargle-periodogram} is the addition
of a multiplicative weight $1/\sigma_n$ within each of the summations.
Early versions of this sort of modification appeared in \citet{Gilliland87} and
\citet{Irwin89}; \citet{Scargle89} derived this ``weighted'' form of the
periodogram without direct reference to the least squares model,
and \citet{Zechmeister09} showed that such a modification does not change
the statistical properties of the resulting periodogram.

This generalization of $\chi^2(f)$ also suggests a convenient way to
construct a periodogram in the presence of correlated observational noise.
If we let $\Sigma$ denote the  $N\times N$ noise
covariance matrix for $N$ observations, and construct the
vectors
\begin{eqnarray}
  &\vec{y} = [y_1, y_2,\cdots y_n]^T \nonumber\\
  &\vec{y}_{model} = [y_{model}(t_1),y_{model}(t_1),\cdots y_{model}(t_n)]^T,
\end{eqnarray}
then the $\chi^2$ expression for correlated errors can be written
\begin{equation}
  \chi^2(f) = (\vec{y}-\vec{y}_{model})^T\Sigma^{-1}(\vec{y}-\vec{y}_{model}),
  \eqlabel{chi2-with-correlated-errors}
\end{equation}
which reduces to \eq{chi2-with-errors} if noise is uncorrelated (i.e., if the
off-diagonal terms of $\Sigma$ are zero).
This resulting periodogram is quite similar to the approach to correlated
noise developed by \citet{Vio2010} from the Fourier perspective,
and is in fact exactly equivalent in the case of the
``zero-mean colored noise'' example considered therein.

\subsection{Data Centering and the Floating Mean Periodogram}
\sectlabel{floating-mean}

Another commonly-applied modification of the periodogram has variously been
called the
{\it Date-compensated Discrete Fourier Transform} \citep{Ferraz-Mello81},
the {\it floating-mean periodogram} \citep{Cumming99,VanderPlas2015},
or the {\it generalized Lomb-Scargle Method} \citep{Zechmeister09},
and involves adding an offset term to the sinusoidal model at each
frequency\footnote{We choose to follow \citet{Cumming99} and
  \citet{VanderPlas2015} and call this a ``Floating Mean'' model,
  to avoid confusion of the different uses of the term
  ``generalized periodogram'' in, {e.g.} \citet{Bretthorst2001} and
  \citet{Zechmeister09}.
}:
\begin{equation}
  y_{model}(t;f) = y_0(f) + A_f \sin(2 \pi f (t - \phi_f))
  \eqlabel{floating-mean-model}
\end{equation}
This turns out to be quite important in practice, because the standard
Lomb-Scargle approach assumes that the
data are pre-centered around the mean value of the (unknown) signal.
In many analyses, this requirement is accomplished by pre-centering data
about the sample mean: this approach is generally robust when
the data provide full phase coverage of the observed signal;
however, due to selection effects and survey cadence, full phase
coverage can not always be guaranteed.
Using the sample mean in such cases can potentially lead to suppression of
peaks of interest.

\begin{figure}[ht]
  \centering
  \includegraphics[width=\textwidth]{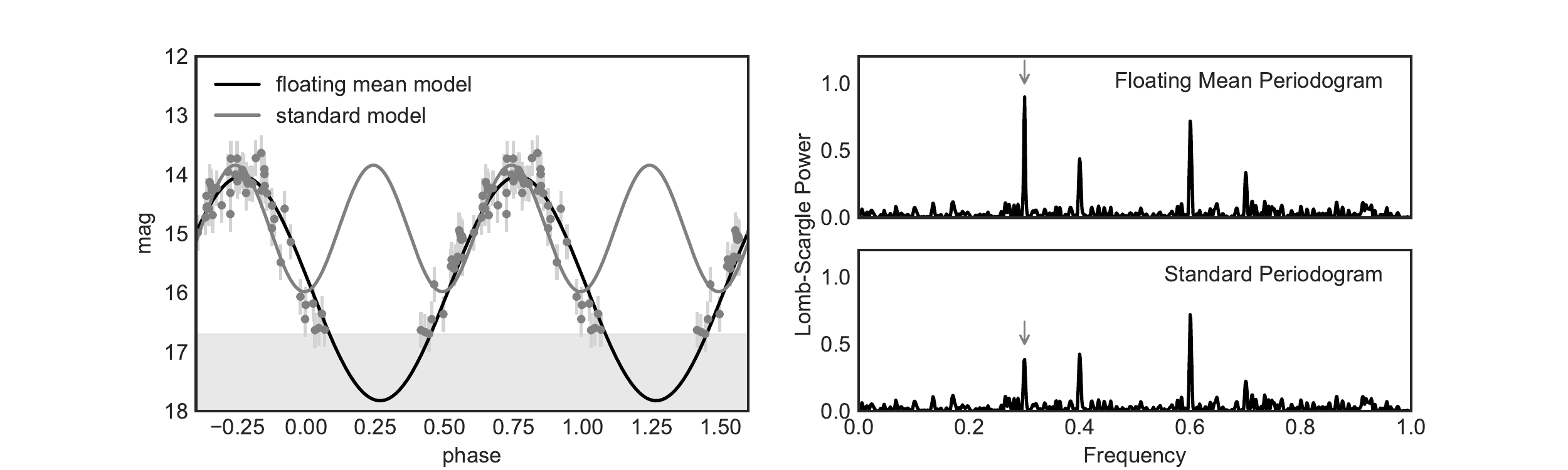}
  \caption{A comparison of the standard and floating mean periodograms for
    data with a frequency of 0.3 and a selection effect which removes
    faint observations.
    In this case the mean estimated from the observed data is not close to
    the true mean, which leads to the failure of the standard periodogram to
    recover the correct frequency. A floating mean model correctly recovers
    the true frequency of 0.3.
    \figlabel{standard-vs-floatingmean}}
\end{figure}

A simulated example of this is shown in \Fig{standard-vs-floatingmean}: the data
consist of noisy observations of a sinusoidal signal in which the faintest
observations are omitted from the dataset (due to, e.g., a detection threshold).
Applying the standard Lomb-Scargle periodogram to pre-centered data leads to a
periodogram that suppresses the true period of 0.3 days (lower right panel).
Using the floating-mean model of \eq{floating-mean-model} yields a periodogram
that identifies this true period (upper right panel).
A detailed study of the floating-mean model is given by
\citet{Zechmeister09} who show that the addition of the floating mean term
does not change the useful statistical properties outlined in \sect{schuster-to-lomb-scargle}.

\subsection{Higher-order Fourier Models}
\sectlabel{multiterm}

A further generalization of the least squares periodogram involves multi-term Fourier models:
rather than fitting just a single sinusoid at each frequency, we might fit a
partial Fourier series, adding $K-1$ additional sinusoidal components at
integer multiples of the fundamental frequency:
\begin{equation}
  y_{model}(t;f) = A_f^{0} + \sum_{k=1}^K A_f^{(k)} \sin(2\pi k f (t - \phi_f^{(k)})).
\end{equation}
\citet{Bretthorst88} takes a comprehensive look at this type of multi-term
extension to the periodogram, as well as related extensions to decaying signals
and other more complex models.

\begin{figure}[ht]
  \centering
  \includegraphics[width=\textwidth]{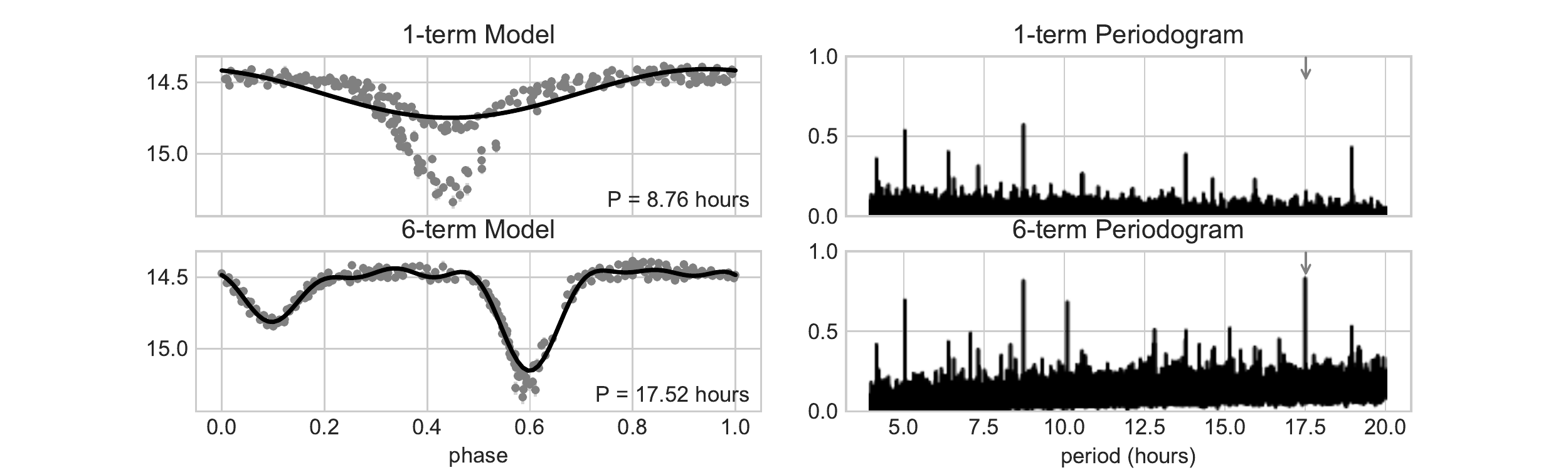}
  \caption{1-term and 6-term Lomb-Scargle models fit to LINEAR object
    14752041, an eclipsing binary. Notice that the standard periodogram
    (upper panels) finds an alias of the true 17.5-hour frequency,
    because a simple sinusoidal model cannot closely fit both the primary
    and secondary eclipse. A six-term Fourier model, on
    the other hand, does find the true period (lower panels).
    \figlabel{binary-multiterm}}
\end{figure}

This kind of Fourier series generalization to the periodogram is quite
tempting, because it means that the periodogram can be tuned to fit models
that are more complicated than simple sine waves.
In some cases, this can be very useful; for example, \fig{binary-multiterm}
shows a Lomb-Scargle analysis of an eclipsing binary star, characterized by
both a primary and secondary eclipse.
The standard Lomb-Scargle periodogram (upper panels) is maximized at twice the true rotation
frequency, because the simple sinusoidal model is unable to closely fit the
primary and secondary eclipses separately.
A six-term Fourier model (lower panels) is sufficiently flexible that it can
detect the true 17.5-hour period, though this comes at the expense of a much
noisier periodogram.

This additional periodogram noise is easy to understand: in the least-squares
view of the periodogram, the periodogram height at any frequency is directly
related to how well the model fits the data.
For nested linear models, adding additional terms will {\it always} provide an
equal or better fit to data than the simpler model, and so it follows that
a periodogram based on a more complex model will be higher
{\it at all frequencies}, not just at frequencies of interest!
Indeed, this effect can be readily observed in \fig{binary-multiterm}.

\begin{figure}[ht]
  \centering
  \includegraphics[width=\textwidth]{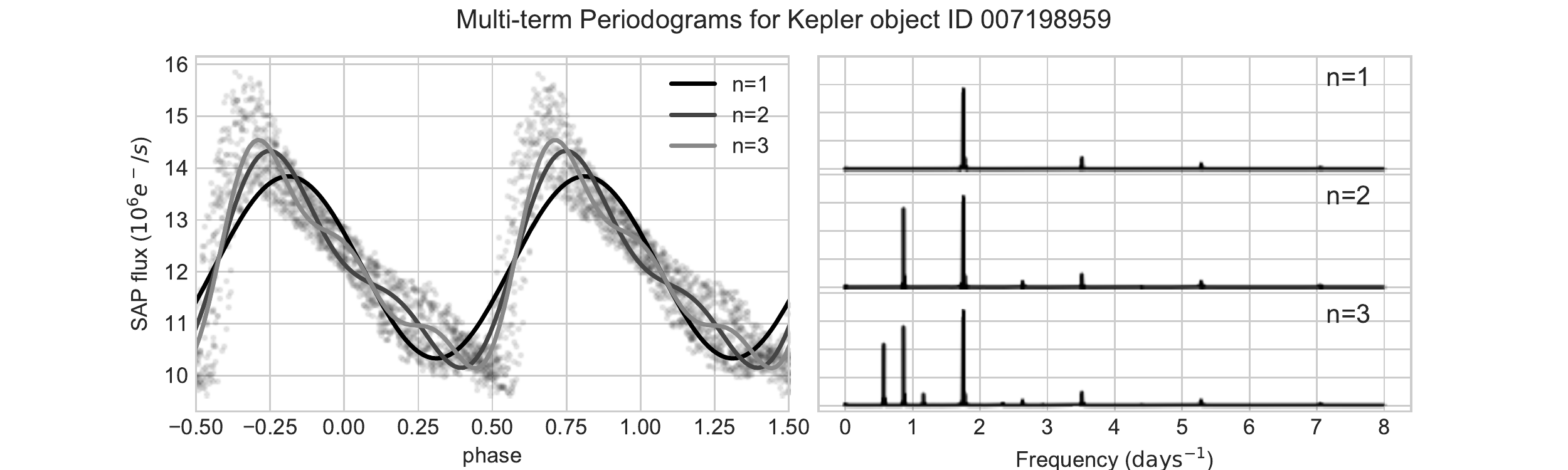}
  \caption{Multi-term models (left) and their corresponding periodograms (right)
    for the kepler data shown in \fig{kepler-data}
    \figlabel{kepler-multiterm}}
\end{figure}

This background noise in the multi-term periodogram can also be understood
in terms of aliasing.
Consider \fig{kepler-multiterm}, which shows several multi-term fits to the
Kepler data from \fig{kepler-data}.
Given a standard periodogram with a peak or sub-peak at $f_0$,
a 2-term periodogram will duplicate this peak at $f_0 / 2$,
with the second harmonic driving the fit.
Similarly, a 3-term periodogram will add additional peaks both at
$f_0 / 2$ and $f_0 / 3$, due to the original peak falling in the second and
third harmonic.
In general, you can expect an $N$-term periodogram to contain $N$ aliases of
every feature present in the standard periodogram; any strong peak revealed
by the multi-term periodogram is due to two or more of these aliases coinciding.

\begin{figure}[ht]
  \centering
  \includegraphics[width=\textwidth]{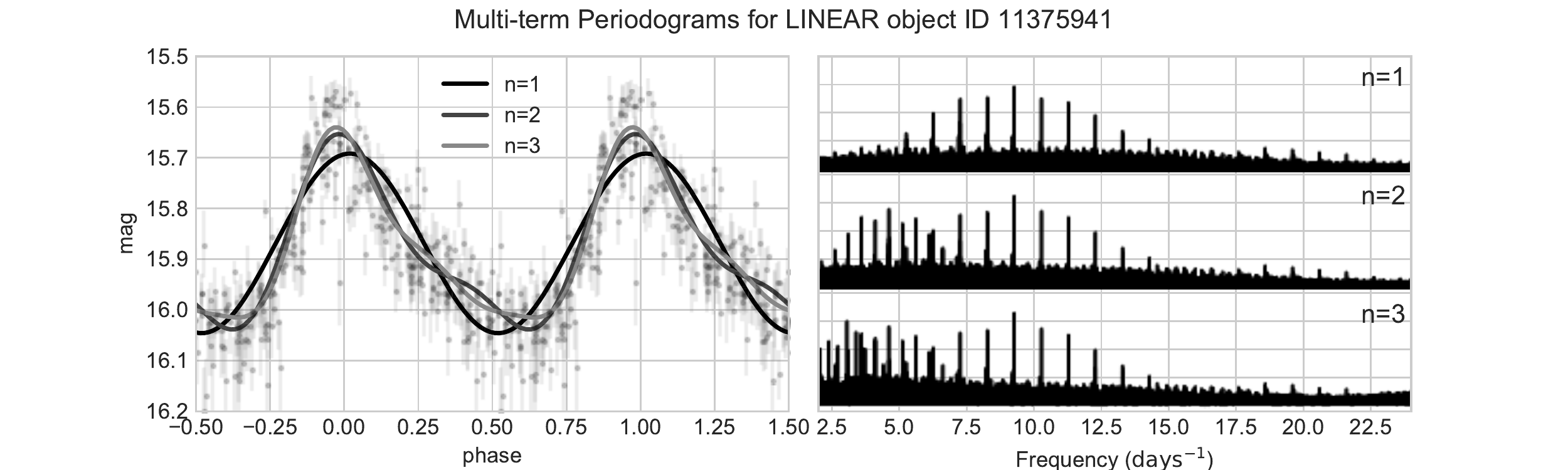}
  \caption{Multi-term models (left) and their corresponding periodograms (right)
    for the LINEAR data shown in \fig{LINEAR-data}
    \figlabel{LINEAR-multiterm}}
\end{figure}

In cases where the single-term power spectrum is itself noisy and/or dominated
by partial aliasing due to window effects, these additional aliases can
quickly wash-out any gains from the more complicated model.
For example, consider the multi-term periodograms for the LINEAR data
depicted in \fig{LINEAR-multiterm}: as previously discussed,
the LINEAR periodogram contains a number of aliases of the true peak due
to the dominant 1-day signal in the window function.
Adding terms to this model only compounds this problem, increasing the number of
spurious peaks at the low-frequency end.
For data with even moderate levels of noise, chance coincidences of these
peaks can lead to spurious detections which dominate the true peak, particularly
for models with many Fourier terms.

\subsection{Additional Extensions}
When the periodogram is viewed from the least-squares model-fitting perspective,
there is no need to limit the analysis to sums of sinusoids.
There have been some very interesting applications extending this type of
analysis to more arbitrary models.
For example, periodogram models can be extended to measure
decaying signals, non-stationary signals, multi-frequency signals, chirps,
and other signal types \citep[see, e.g.][]{Jaynes87, Bretthorst88, Gregory2001}.
In particular, \citet{Bretthorst88} demonstrates the effectiveness of such
approaches in applications ranging from medical imaging to astronomy.
The challenge of such extensions is the fact that you often need to use some
prior knowledge of the system being observed to decide whether a more
complicated model is indicated, as well as what form of model to apply.
In practice this comes up only when searching for very specific signals for
which a more complicated model has some {\it a priori} physical motivation.

On the astronomy side, several examples of this flavor of extension exist.
For example, the Supersmoother approach to detecting periodicity
involves fitting a flexible non-parametric smoothing
function to the data at each frequency \citep{Reimann94}: the flexibility
of the model leads to fewer aliasing issues when compared to the more
constrained sinusoidal model.
Another approach is to use empirically-derived templates as a functional fit
at each frequency; this has been employed effectively in \citet{Sesar2010,
Sesar2013} and related work.

Finally, there has been some exploration of extensions to the Lomb-Scargle
periodogram for use with multi-band observations, using various forms of
regularization to control model complexity \citep{VanderPlas2015, Long2016}.
With many of these least-squares and/or Bayesian extensions,
computational complexity quickly becomes an issue, because fast
$\mathcal{O}(N\log N)$ approaches which can be used for sinusoidal fits
(see \sect{algorithmic-considerations}) are not available for more general
functional forms, though there is some promising work in this area: see,
for example, the {\it Fast Template Periodogram}\footnote{\url{http://ascl.net/code/v/1559}}
(Hoffman et al. 2017, {\it in prep})
which can quickly construct periodograms from Fourier approximations to templates.

\subsection{A Note About ``Bayesian'' Periodograms}
\sectlabel{bayesian-periodograms}
The least squares view of the Lomb-Scargle periodogram creates a natural
bridge, via maximum likelihood, to Bayesian periodic analysis.
In fact, \citet{Jaynes87} showed that the standard form of the Lomb-Scargle
periodogram can be derived directly from the axioms of Bayesian probability
theory outlined in his comprehensive treatment of the subject
\citep{Jaynes03}\footnote{Read this. Really.}.
In the Bayesian view, the Lomb-Scargle periodogram is in fact the optimal
statistic for detecting a stationary sinusoidal signal in the presence of
Gaussian noise.

For the standard, simple-sinusoid model, the Bayesian periodogram is given by the posterior
probability of frequency $f$ given the data $D$ and sinusoidal model $M$:
\begin{equation}
  p(f\mid D, M) \propto e^{P_{LS}(f)}
  \eqlabel{bayesian-periodogram}
\end{equation}
where $P_{LS}(f)$ is the Lomb-Scargle power from \eq{lomb-scargle-periodogram}.
The effect of this exponentiation, as seen in \fig{bayesian-periodogram},
is to suppress side-lobes and alias peaks in relation to the largest one
of the spectrum.

\begin{figure}[ht]
  \centering
  \includegraphics[width=\textwidth]{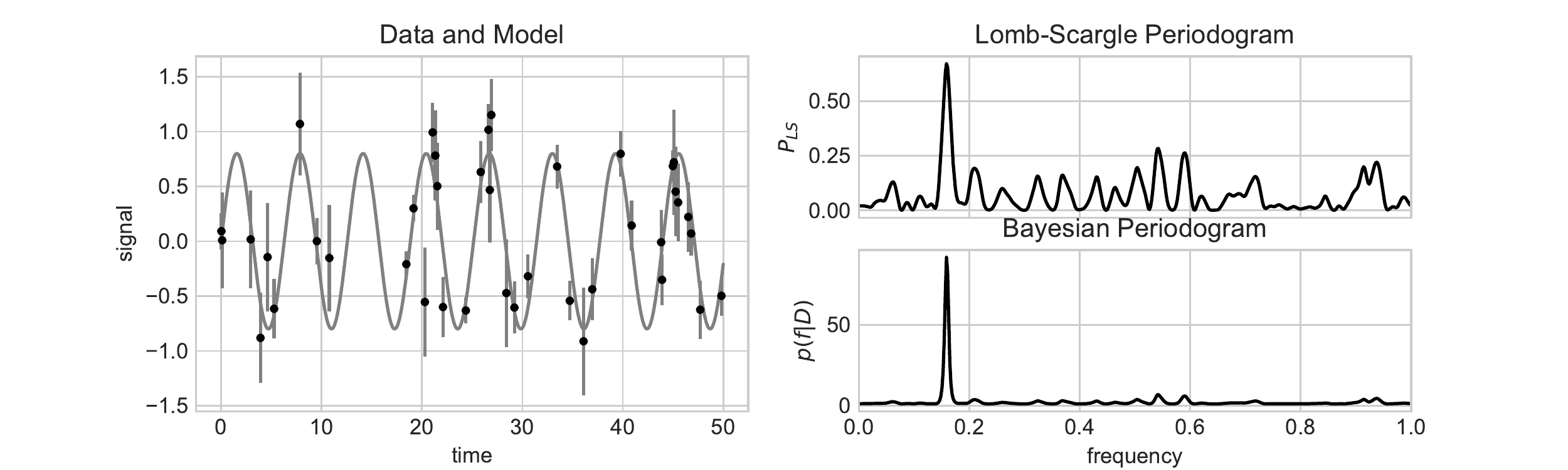}
  \caption{Comparison of the Lomb-Scargle periodogram (upper right) and the
    Bayesian posterior periodogram (lower right) for simulated data (left)
    drawn from a simple sinusoid.
    The Bayesian posterior is equal to the exponentiation of the unnormalized
    Lomb-Scargle periodogram, and thus tends to suppress all but the largest
    peak. The Bayesian approach can be useful, but is problematic if
    not used carefully (see text for discussion).
    \figlabel{bayesian-periodogram}}
\end{figure}

The benefit of the Bayesian approach is that it allows more flexible models,
more principled treatment of nuisance parameters related to noise in the data,
and the ability to make use of prior information in a periodic analysis.
It explicitly returns probability density as a function of frequency,
and it is tempting to think of this as the
``natural'' way to interpret the periodogram.

The problem, however, is that the Bayesian periodogram does {\it not} compute
the probability that the data are periodic with a given frequency, but rather
that probability conditioned on a relatively strong assumption:
that {\it the data are drawn from a sinusoidal model}.
As such, the standard Bayesian periodogram is not a useful measure of
generic periodocity!
In fact, the Baysian periodogram's derivation under the
assumption of a sinusoidal signal is perhaps the
best argument against its use for unknown signals:
the result of a Bayesian analysis is only ever as good as the assumptions that
go into it, and for a general (not-necessarily sinusoidal) signal, those
assumptions are incorrect, and the periodogram should not be trusted.

Now, the standard Lomb-Scargle periodogram can also be viewed as derived
from a sinusoidal fit, and thus might appear subject to the same criticism.
But unlike the Bayesian periodogram, the standard periodogram affords
interpretation in light of Fourier analysis and window functions: here
the incorrect sinusoidal model manifests itself in terms of frequency
aliases, which can be understood through the analysis of window functions.
In most cases of interest, such analysis turns out to be {\it vital} to the
application and interpretation of the periodogram (see \sect{failure-modes}).

In other words, the Bayesian approach essentially
goes ``all-in'' on the least squares
interpretation of the periodogram, exponentially suppressing the information
that allows you to reason about the periodogram in light of its connection
to Fourier analysis.
In contrived cases with clean sinusoidal data and unstructured window
functions, exponential attenuation
of side-lobes and aliases may seem appealing \citep[see, e.g.][]{Mortier15},
but that appeal extends to the real world only in the most favorable of
cases---i.e., high signal-to-noise measurements of near-sinusoidal data with
a very well-behaved survey window.
In short, you shiould be wary of placing too much trust in a Bayesian
periodogram, unless you're certain of the type of signal you're looking for.

This is not to say that all Bayesian approaches to periodic analysis are
similarly flawed; there have been many interesting studies that go beyond
the simple sinusoidal model and use more complex and/or flexible models.
Some examples are models based on Fourier
extensions with strong priors \citep[e.g.][]{Bretthorst88},
instrument-dependent parametric models \citep[e.g.][]{Angus16},
flexible non-parametric functions \citep[e.g.][]{Gregory92},
Gaussian Process models \citep[e.g.][]{Wang2012}, and
specially-designed stochastic models \citep[e.g.][]{Kelly14}.
Though these are often be more accurate and powerful than the Lomb-Scargle
approach, they tend to be far more expensive computationally.
Bayesian approaches based on Markov Chain Monte Carlo (MCMC) also tend to
run into stability problems,
particularly for multimodal or other complicated posterior distributions
\citep[See, for example, the RR Lyrae discussion in][]{Kelly14}.

\section{Practical Considerations when using Lomb-Scargle Periodograms}
\sectlabel{practical-considerations}
The previous sections have given a conceptual introduction to the Lomb-Scargle
periodogram and its roots in both Fourier and Least-squares analysis.
This section gets to the meat of the subject at hand:
given this understanding of the Lomb-Scargle periodogram and related
approaches, how should we use it most effectively in practice?
The following sections will identify several of the important issues and
questions that are not often addressed in the literature on the subject,
but are nevertheless vital to consider when using the periodogram in practice.

\subsection{Choosing a Frequency Grid}
\sectlabel{frequency-grid}
The frequency grid used for a Lomb-Scargle analysis is an important choice
that is too-often glossed-over, probably because the choice is so
straightforward in the more familiar case of uniformly-sampled data.
For non-uniform data, it is not so simple, and there are two important
considerations: the {\it frequency limits} and the {\it grid spacing}.

The frequency limit on the low end is relatively easy: for a set of observations
spanning a length of time $T$,
a signal with frequency $1/T$ will complete exactly one oscillation cycle,
and so this is a suitable minimum frequency.
Often, it's more convenient just to set this minimum frequency to zero, as it
doesn't add much of a computational burden and is unlikely to add any
significant spurious peak to the periodogram.

The high-frequency limit is more interesting, and goes back to the discussion
of Nyquist and/or limiting frequencies from \sect{pseudo-nyquist}:
in order to not miss relevant information, it is important to compute the
periodogram up to some well-motivated limiting frequency $f_{max}$. This could
be a true Nyquist limit based on the \citet{Eyer99} definition, a pseudo-Nyquist
limit based on careful scrutiny of the window function (cf. \fig{kepler-data}),
a limiting frequency based on the integration time of individual
observations, or a limit based on prior knowledge of the kinds of signals
you expect to detect.

\begin{figure}[ht]
  \centering
  \includegraphics[width=\textwidth]{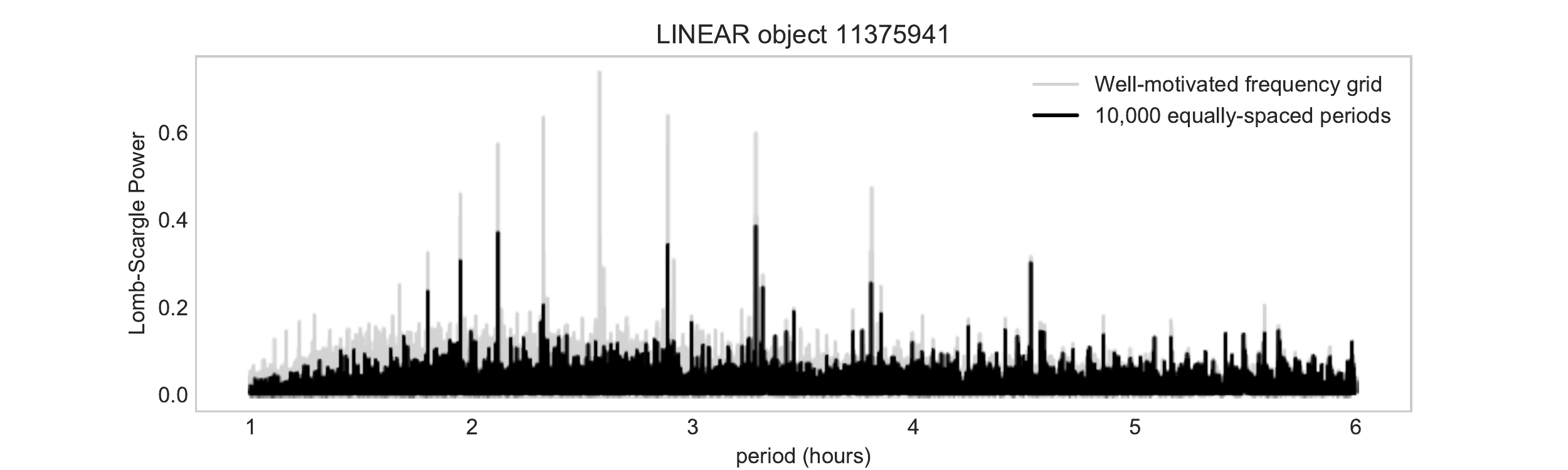}
  \caption{An example of a poorly-chosen frequency grid for the data in
    \fig{LINEAR-data}. The dark curve shows a periodogram evaluated on a
    grid of 10,000 equally-spaced periods; the light curve shows the true
    periodogram (evaluated at ${\sim}200,000$ equally-spaced frequencies).
    This demonstrates that using too coarse a grid can lead to a periodogram
    that entirely misses relevant peaks.
    \figlabel{LINEAR-coarse-grid}}
\end{figure}

With the frequency range decided, we next must determine how finely to sample
the frequencies between the limits.
This choice turns out to be quite important as well: too fine a grid can
lead to unnecessarily long computation times that can add up quickly in the
case of large surveys, while too coarse a grid risks
entirely missing narrow peaks that fall between grid points.
For example, \fig{LINEAR-coarse-grid} shows the true, well-sampled periodogram
(gray line), along with a periodogram computed for 10,000 equally-spaced
periods covering the same range (black line).
Because the spacing of the 10,000-point grid is much larger than the width of
the periodogram peaks, the analysis {\it entirely misses} the most
important peaks in the periodogram!

This shows us that it's important to choose grid spacings smaller than the
expected widths of the periodogram peaks.
From our discussion of windowing in \sect{window-functions}
(particularly \fig{rectangular-window}), we know that data observed through a
rectangular window of length $T$ will have sinc-shaped peaks of width
${\sim}1/T$.
To ensure that our grid sufficiently samples each peak, it is prudent to
over-sample by some factor---say $n_o$ samples per peak---and
use a grid of size $\Delta f = \frac{1}{n_o T}$.
This pushes the total number of required periodogram evaluations to
\begin{equation}
  \eqlabel{n-eval}
  N_{eval} = n_o T f_{max}
\end{equation}
So what is a good choice for $n_o$?
Values ranging from $n_o=5$ \citep{Schwarzenberg-Czerny96, VanderPlas2015}
to $n_o=10$ \citep{Debosscher07, Richards12}
seem to be common in the literature;
for periodograms computed in this paper, we use $n_o=5$.
% \citep{Schwarzenberg-Czerny96} discusses how $n_o$ can be selected
% more carefully based on factors such as oversampling, number of harmonics
% in the model, and other considerations.

Depending on the characteristics of the dataset, the size of the resulting
frequency grid can vary greatly.
For example, the Kepler data shown in \fig{kepler-data} has a pseudo-Nyquist
frequency of $48.9\ {\rm days}^{-1}$ and
an observing window of $T = 90\ {\rm days}$.
To compute five samples per peak thus requires $N_{eval} \approx 22,000$
evaluations of the periodogram.
On the other hand, the LINEAR data shown in \fig{LINEAR-data} does not have
any noteable aliasing structure in its window function.
In this case the maximum detectable frequency is the Nyquist limit defined
by its temporal resolution, which is 6 digits beyond the decimal point in days.
From \eq{nonuniform-nyquist}, we can write $f_{Ny} = 500,000\ {\rm days}^{-1}$,
and given the observing window of $T = 1962\ {\rm days}$, we find that five
evaluations per peak across the entire detectable frequency range
would require $N_{eval} \approx 4.9 \times 10^9$ evaluations of the periodogram!
Computing this large a periodogram in most cases is computationally intractable
(see \sect{algorithmic-considerations}), and so in practice one must choose
a smaller limiting frequency based on prior information about what kind of
signals are expected in the data: for example, in \fig{LINEAR-power}, we chose
a limiting frequency of $f_{max} = (1\ {\rm hour})^{-1}$ based on typical
oscillation periods expected for SX Phe-type stars.
This leads to a much more manageable
$N_{eval} \approx 240,000$ periodogram evaluations.

By comparison, data from the LSST survey \citep{Ivezic08LSST}
will fall somewhere in-between: full frequency coverage of
the 10-year data up to a limiting frequency defined by the 30-second
integration time for each visit would require roughly 25 million
periodogram evaluations per object, which for fast implementations
(see the next section) could be accomplished in several seconds on
a modern CPU.

One final note: although it can be more easily interpretable to
visualize periodograms as a function of period rather than a function of
frequency, the peak widths are not constant in period.
Regular grids in period rather than frequency tend to over-sample
at large periods and under-sample at small periods;
for this reason it is preferable to use a regular grid in frequency.

\subsection{Failure Modes}
\sectlabel{failure-modes}

When using the Lomb-Scargle periodogram in an observational pipeline, it is
vital to keep in mind the failure modes of the periodogram approach, which
are rooted in the aliasing and pseudo-aliasing effects rooted in the structure
of the window function (recall \sect{window-functions}).
Due to the interaction of the signal, the convolution due to the survey window,
and noise in the data, it is quite common for the largest peak in the
periodogram to correspond not to the true frequency, but some alias of that
frequency.

\begin{figure}[ht]
  \centering
  \includegraphics[width=\textwidth]{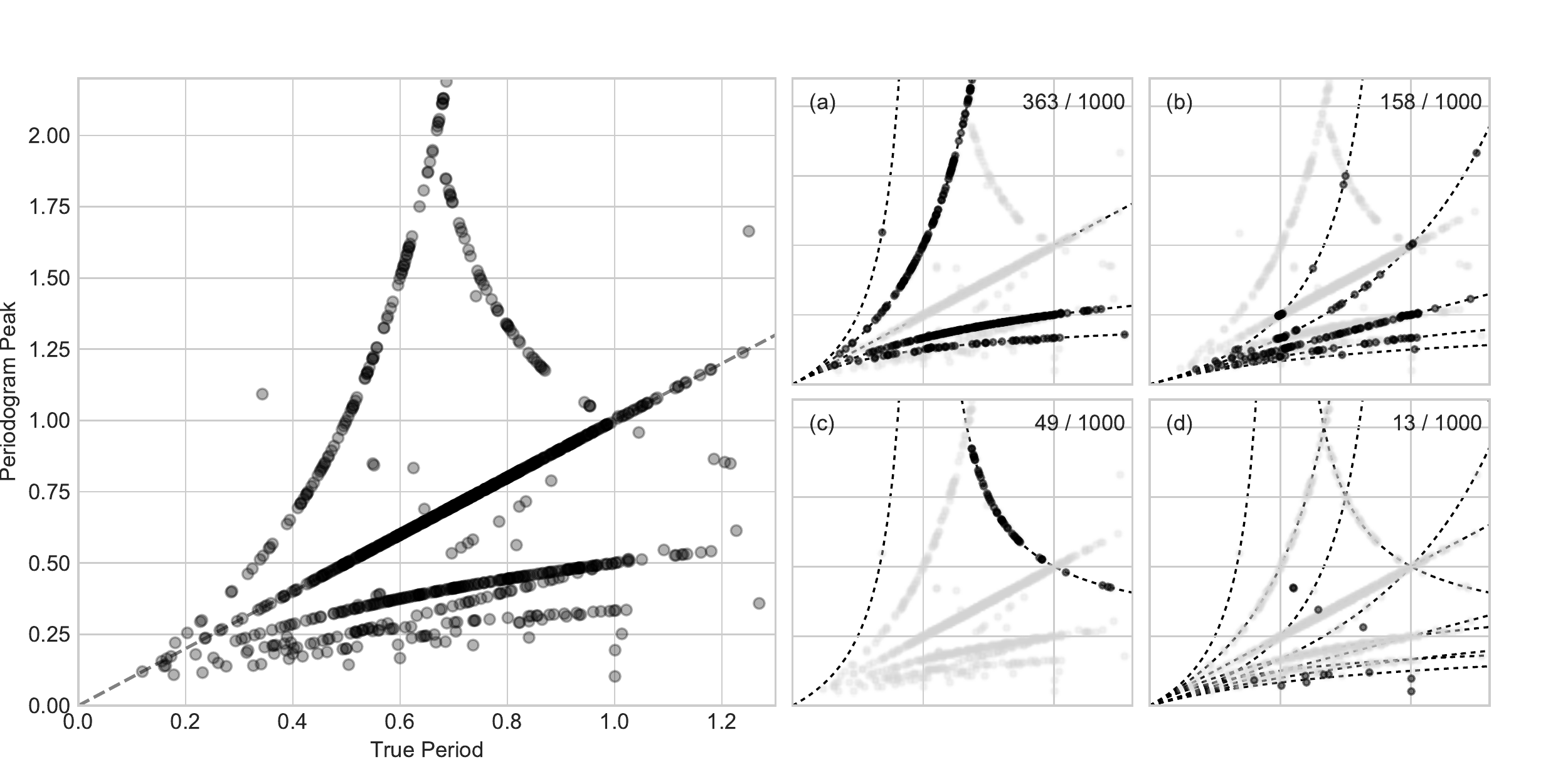}
  \caption{Comparison of the true period and peak Lomb-Scargle period for
    1000 simulated periodic light curves.
    Each has 60 irregular observations over 180 nights,
    with a cadence typical of ground-based surveys (i.e., showing a strong
    diurnal window pattern similar to \fig{LINEAR-window}).
    The Lomb-Scargle peak does not always coincide with the true period,
    and there is noticeable structure among these failures.
    Panels (a)-(d) isolate some of the specific modes of failure that should
    be expected for this kind of window function;
    see the text for more discussion.
    \figlabel{failure-modes}}
\end{figure}

\fig{failure-modes} demonstrates this for some simulated data.
The data consist of 60 noisy observations each of 1,000 simulated light
curves within a span of 180 days.
The window is typical of ground-based data, with each observation recorded
within a few hours before or after midnight each night.
The left panel shows the results: the peridogram peak coincides with the true
period in under 50\% of cases, and the modes of failure lead to noticeable
patterns in the resulting plot.
Though these are simulated observations, the pattern seen here is typical of
real observations: see, for example, \citet{VanderPlas2015}
and \citet{Long2016} which show similar plots for RR Lyrae candidates from
the Sloan Digital Sky Survey.

These patterns can be understood in terms of the interaction between the
window function and the underlying spectral power.
As we discussed in \sect{window-functions},
a nightly observation pattern---typical of ground-based surveys---leads to
a window function with a strong diurnal component that causes each frequency
signature $f_0$ to be partially aliased at $f_0 + n \delta f$, for integers
$n$ and $\delta f = 1$ cycle/day.
(Recall \figs{LINEAR-window}{LINEAR-window-effect}).
This is the source of the failure modes depicted in panel (a) of
\fig{failure-modes}; in terms of periods the above expression becomes:
\begin{equation}
  P_{obs} = \left(\frac{1}{P_{true}} + n\ \delta f\right)^{-1}
  \eqlabel{alias-periods}
\end{equation}
where, to be clear, $n$ is a positive or negative integer and $\delta f$ is
the frequency of a strong feature in the window  (here, $1$ cycle/day).
For the simulated data, close to 36\% of the objects are mis-characterized
along these failure modes.

Another mode of failure is when the periodogram for an object with frequency
$f_0$ isolates a higher harmonic of that fundamental frequency, such as $2f_0$.
This may occur for periodic signals that are not strictly sinusoidal, and so
have power at a higher harmonics.
This higher harmonic peak is also subject to the same aliasing
effects as in \eq{alias-periods}.
Thus we can extend \eq{alias-periods} and describe the failure modes depicted
in panel (b) of \fig{failure-modes} with the following equation,
for positive integers $m > 0$:
\begin{equation}
  P_{obs} = \left(\frac{m}{P_{true}} + n\delta f\right)^{-1}
  \eqlabel{alias-harmonic-periods}
\end{equation}
Panel (b) of \fig{failure-modes} illustrates this for $m=2$; this harmonic and
its aliases account for roughly 15\% of the results.

Panel (c) of \fig{failure-modes} shows the final pattern of failure in
Lomb-Scargle results, which has an opposite trend with true period.
This is closely related to the effects shown in panels (a) and (b),
but comes from the even symmetry of the Lomb-Scargle periodogram.
Every periodogram peak at frequency $f_0$ has a corresponding peak at $-f_0$,
and this is true of aliases as well.
If a peak's aliases cross into the negative frequency regime, they are
effectively reflected into the positive-frequency range.
This reflection can be accounted for by a further modification of
\eq{alias-harmonic-periods}---taking its absolute value:
\begin{equation}
  P_{obs} = \left|\frac{m}{P_{true}} + n\delta f\right|^{-1}
  \eqlabel{failure-modes}
\end{equation}
Panel (c) shows the 5\% of objects that fall along this reflected
failure mode for $m=1$, $n=-2$.
After accounting for all these known sources of periodogram failure,
only roughly 1\% of points are misclassified in an ``unexplained'' way,
seen in panel (d).

When applying a periodogram in practice, it is vital to take such effects into
account, rather than blindly relying on the single periodogram peak as your
best estimate of the period.
Applying understanding of windowing and aliasing effects can help in detecting
failures of the periodogram, but is no silver bullet.
For an observed peak at $f_{peak}$ from a survey whose window has strong power
at $\delta f$, something like the following should probably be employed:
\begin{enumerate}
  \item Check for a peak at $f_{peak}/m$ for at least $m \in \{2, 3\}$.
    If a significant peak is found, then $f_{peak}$ is probably an order-$m$
    harmonic of the true frequency.\footnote{Notice that this step can detect
      aliases like those encoundered in \fig{binary-multiterm},
      without resorting to a problematic multiterm model.}
  \item Check for peaks at $|f_{peak} \pm n\delta f|$ for at least
    $n \in \{1, 2\}$ (where $\delta f$ is determined from plotting the survey
    window power, and is generally $(1\ {\rm day})^{-1}$ for ground-based surveys).
    If these aliases exist, then it is possible that you have
    found a peak on the sequence of expected aliases---though keep in mind that
    there is no way to know from the periodogram alone whether or not
    this is the ``true'' peak!
  \item For each of the top few of these aliases, fit a more complicated model
    (such as a multi-term Fourier series, template-based fit, etc.) to select
    among them.
\end{enumerate}
For noisy observations, this procedure cannot generally guarantee that you
have found the correct peak, but it is far preferable to the simplistic
approach of blindly trusting the highest peak in the periodogram!
We will come back to the question of uncertainty in the periodogram result
in \sect{uncertainties}.

\subsection{Window Functions and Deconvolution}
\sectlabel{windows-and-deconvolution}

Our discussion of window functions here and in \sect{window-functions}
has highlighted the impact of structure in the survey window on the
resulting observed power spectrum, and the importance of examining that
window power to understand features of the resulting periodogram.
Here we will consider the computation of the window function, as well
as the possibility of recovering the true periodogram by deconvolving
the window.

\subsubsection{Computing the Window Function}
\sectlabel{computing-window-function}
The window power spectrum can be computed directly from the
delta-function representation of the window function;
From \eq{FT-def}, \eq{power-spectrum}, and \eq{nonuniform-window},
we can write
\begin{equation}
  \mathcal{P}_W(f;\{t_n\}) = \left|\sum_{n=1}^{N} e^{-2\pi i f t_n}\right|^2
  \eqlabel{window-power-analytic}
\end{equation}
Comparing this to \eq{schuster-periodogram}, we see that this is essentially
the classical periodogram for data $g_n=1$ at all times $t_n$.
With this fact in mind, one convenient way to estimate the form of the
window power is to compute a standard Lomb-Scargle periodogram on a series of
unit measurements, making sure to {\it not} pre-center the data or to use a
floating mean model.
As \citet{Scargle82} notes, this computation does {\it not} give the true
window---it differs
from the true window just as the Lomb-Scargle periodogram differs from the
classical periodogram---but in practice it is accurate enough to give
useful insight into the window function's important features.
This method is how window power spectra has been computed throughout this
paper.

\subsubsection{Deconvolution and CLEANing}
\sectlabel{CLEAN}
With this ability to compute the window function, we might hope to be able to
use it {\it quantitatively} to remove spurious peaks from the observed
power spectrum.
Recall from \eq{g-nonuniform} that the observed data are a point-wise product
of the underlying function and the survey window:
\begin{equation}
  g_{obs}(t) = g(t)W_{\{t_n\}}(t),
\end{equation}
and the convolution theorem tells us that the observed Fourier transform is
a convolution of the true transform and the window transform:
\begin{equation}
  \mathcal{F}\{g_{obs}\} = \mathcal{F}\{g\} \ast \mathcal{F}\{W_{\{t_n\}}\}
\end{equation}
Given this relationship, we might hope to be able to invert this
convolution to recover $\mathcal{F}\{g\}$ directly.
For example, we could write:
\begin{eqnarray}
  g(t) &=& g_{obs}(t) / W_{\{t_n\}}(t)\nonumber\\
  \mathcal{F}\{g\} &=& \mathcal{F}\{g_{obs}/W_{\{t_n\}}\}\nonumber\\
                   &=& \mathcal{F}\{g_{obs}\} \ast \mathcal{F}\{1/W_{\{t_n\}}\}
\end{eqnarray}
Because of the localization of observations, $W_{\{t_n\}}(t)$ is zero for most
values of $t$ and so $1/W_{\{t_n\}}(t)$ and its Fourier transform are not well
defined. Because of this, direct deconvolution is not an option in most
cases of interest---in other words, the deconvolution problem is ill-posed
and does not have a unique solution.

There have been a few attempts in the literature to use procedural algorithms
to solve this under-constrained deconvolution problem, perhaps most notably
by adapting the iterative CLEAN algorithm developed for deconvolution in the
context of radio astronomy \citep{Roberts87}.
For cleaning of Lomb-Scargle periodograms, the CLEAN approach is hindered by
three main deficiencies: first and most importantly, the CLEAN algorithm at each
iteration assumes that the highest peak is the location of the primary signal;
this is not always borne out for realistic observations of faint objects where
the cleaning is most necessary (recall the discussion in \sect{failure-modes}).
Second, the convolution takes place at the level of the Fourier transform
rather than the PSD; trying to clean a PSD directly ignores important phase
information. Third, the CLEAN algorithm assumes a classical FFT analysis:
while the Lomb-Scargle periodogram is equivalent to classical periodogram in the limit of
equally-spaced observations, it is not equivalent in the relevant case of
unequal observations (see \sect{schuster-to-lomb-scargle}),
and so any attempt to apply CLEAN to Lomb-Scargle analysis directly would
be fundamentally flawed even if it were not ill-posed to begin with.

The latter two issues could be remedied by focusing on the non-uniform
Fourier transform and the classical periodogram rather than the Lomb-Scargle
modification, but doing so would jettison the benefits of Lomb-Scargle---i.e.,
its useful statistical properties and extensibility via least squares---and
this would still not address the far more problematic first issue.
My feeling is that there is room for a more principled approach to the
unconstrained deconvolution problem for periodograms derived from non-uniform
fast Fourier transforms (perhaps through some sort of L1/lasso regularization
that imposes assumptions of sparsity on the true periodogram) but to date this
does not appear to have been explored anywhere in the literature.

\subsection{Uncertainties in Periodogram Results}
\sectlabel{uncertainties}
An important aspect of reporting results from the Lomb-Scargle periodogram is
the uncertainty of the estimated period.
In many areas of science we are used to being able to report uncertainties in
terms of errorbars, {e.g.} ``the period is $16.3 \pm 0.6$ hours''.
For periods derived from the Lomb-Scargle periodogram, however, uncertainties
generally cannot be meaningfully expressed in this way:
as we saw in the discussion of failure modes in \sect{failure-modes},
the concern for periodograms is more often a disjointed inaccuracy associated
with false peaks and aliases, rather than a more smooth imprecision in
location of a particular peak.

\subsubsection{Peak Width and Frequency Precision}

A periodic signal will be reflected in the periodogram by the presence of a
peak of a certain width and height.
In the Fourier view, the precision with which a peak's frequency can be
identified is directly related to the width of this peak; often the
half-width at half-maximum $f_{1/2} \approx 1/T$ is used.
This can be formalized more precisely in the least squares interpretation of
the periodogram, in which the inverse of the curvature of the peak is
identified with the uncertainty \citep{ICVG2014}---which in the Bayesian
view amounts to fitting a Gaussian curve to the (exponentiated)
peak \citep{Jaynes87, Bretthorst88}. This introduces first-order
dependence on the number of samples $N$ and their average signal-to-noise
ratio $\Sigma$; the scaling is approximately \citep[see, e.g.][]{Gregory2001}:
\begin{equation}
  \sigma_f \approx f_{1/2} \sqrt{\frac{2}{N\Sigma^2}}.
  \eqlabel{bayes-error}
\end{equation}
This dependence comes from the fact that the Bayesian uncertainty is related
to the width of the {\it exponentiated} periodogram, which
depends on $P_{max}$, the height of the peak\footnote{
To see why, consider a periodogram with maximum value
$P_{max} = P(f_{max})$, so that $P(f_{max} \pm f_{1/2}) = P_{max}/2$.
The Bayesian uncertainty comes from approximating the exponentiated peak as a
Gaussian; i.e., $\exp[P(f_{max} \pm \delta f)] \propto \exp[-\delta f^2/(2\sigma_f^2)]$.
From this we can write $P_{max}/2 \approx P_{max} -f_{1/2}^2 / (2 \sigma_f^2)$ or
$\sigma_f \approx f_{1/2} / \sqrt{P_{max}}$.
In terms of signal-to-noise ratio
$\Sigma = {\rm rms}[(y_n - \mu) / \sigma_n]$,
a well-fit model gives
$P_{max} \approx \hat{\chi}_0^2/2 \approx \Sigma^2 N/2$,
which leads to the expression in \eq{bayes-error}.
}.

We have motivated from the Fourier window discussions that peak width $f_{1/2}$
is the inverse of the observational baseline; the surprising result is that to
first order the peak width in the periodogram {\it does not depend} on
either the number of observations or their signal-to-noise ratio!
This can be seen visually in \fig{peak-width-height},
which shows periodograms for simulated data with a fixed 4-period baseline,
with varying sample sizes and signal-to-noise ratios.
In all cases, the widths of the primary peak are essentially {\it identical},
regardless of the quality or quantity of data!
Data quality and quantity are reflected in the height of the peak in relation
to the ``background noise'', which speaks to the peak {\it significance} rather
than the precision of frequency detection.

\begin{figure}[ht]
  \centering
  \includegraphics[width=\textwidth]{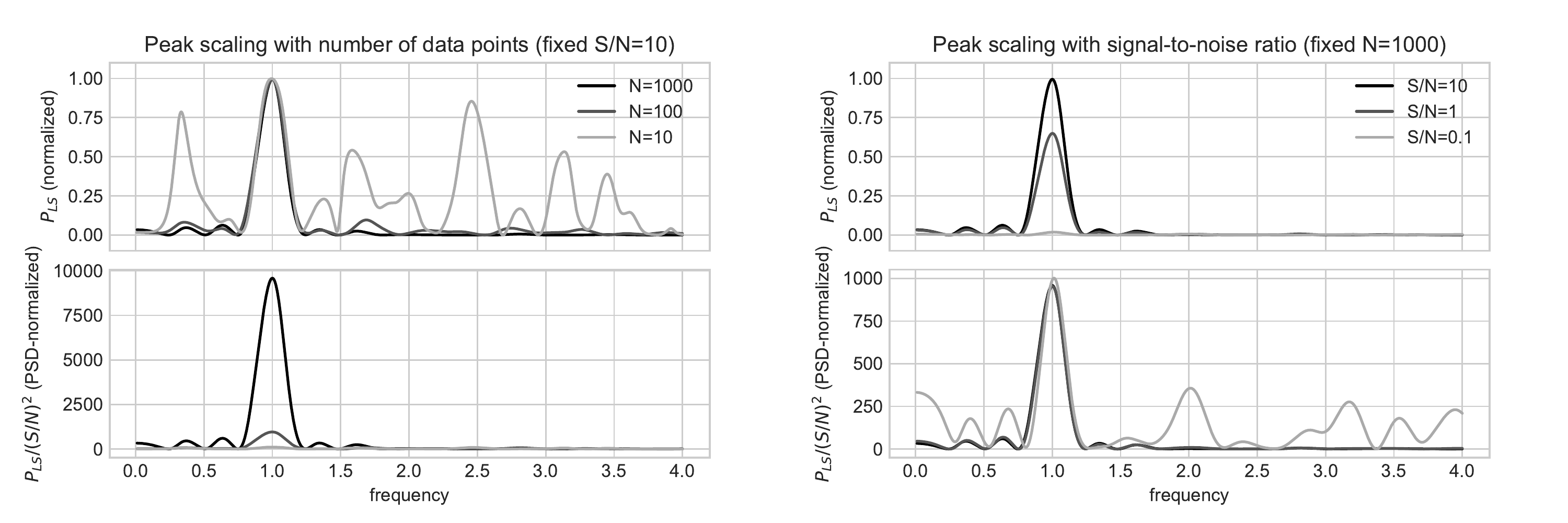}
  \caption{The effect of the number of points $N$ and the signal-to-noise
    ratio $S/N$ on the expected width and height of the periodogram peak.
    Top panels show the normalized periodogram (\eq{lomb-scargle-normalized}),
    while bottom panels show the PSD-normalized periodogram
    (\eq{lomb-scargle-periodogram}) scaled by noise variance.
    Perhaps surprisingly, neither the number of points nor the signal-to-noise
    ratio affects the peak width. \figlabel{peak-width-height}}
\end{figure}

For this reason, if you insist on reporting frequency errorbars derived from
peak width, the results can be pretty silly for observations with
long baselines.
For example, going by the peak width in \fig{LINEAR-power}, the
periodogram reveals a period of $2.58014 \pm 0.00004$ hours, a relative
precision of about 1/1000 percent.
While this is an accurate characterization of the period precision
{\it assuming that peak is the correct one}, it does not capture the
much more relevant uncertainties demonstrated in \sect{failure-modes},
in which we might find ourselves on the wrong peak entirely.
This is why in general, peak widths and Gaussian error bars should generally
be avoided when reporting uncertainties in the context of a periodogram
analysis.

\subsubsection{False Alarm Probability}
\sectlabel{false-alarm-probability}

A much more relevant quantity for expressing uncertainty of periodogram results
is the height of the peak, and in particular the height compared to the
spurious background peaks that arise in the periodogram.
\fig{peak-width-height} indicates that this property {\it does} depend on
both the number of observations and their signal-to-noise ratio: for the
small-$N$/low signal-to-noise cases, the spurious peaks in the background
become comparable to the size of the true peak.
In fact, as we saw in \sect{schuster-to-lomb-scargle}, the ability to
analytically define and quantify the relationship between peak height
and significance is one of the primary considerations that led to the
standard form of the Lomb-Scargle periodogram.

The typical approach to quantifying the significance of a peak is the {\it
False Alarm Probability} (FAP), which measures the probability that a dataset
with no signal would---due to coincidental alignment among the random
errors---lead to a peak of a similar magnitude.
\citet{Scargle82} showed that for data consisting of pure Gaussian noise,
the values of the unnormalized
periodogram in \eq{lomb-scargle-periodogram} follows a $\chi^2$ distribution
with two degrees of freedom; that is, at any given frequeny $f_0$,
if $Z = P(f_0)$ is the periodogram value from \eq{lomb-scargle-periodogram},
then
\begin{equation}
  P_{single}(Z) = 1 - \exp(-Z)
  \eqlabel{LS-prob-z}
\end{equation}
is the cumulative probability of observing a periodogram value less than $Z$,
in data consisting only of Gaussian noise.\footnote{
  Be aware that for different periodogram normalizations
  (See \sect{normalizations}), the form of this distribution changes;
  see \citet{Cumming99} or
  \citet{Baluev2008} for a good summary of the statistical properties of
  various periodogram normalizations.
}

\subsubsubsection{Independent Frequency Method}

We are generally not interested in the distribution of one particular
randomly-chosen frequency,
but rather the distribution of {\it the highest peak} of the
periodogram, which is a quite different situation.

By analogy, consider rolling a standard 6-sided die.
The probability, in a single roll, of rolling a number, say $r > 4$ is easy
to compute: it's 2 sides out of six, or $p(r>4) = 1/3$.
If, on the other hand, you roll 10 dice and choose the largest number among
them, the probability that it will be greater than 4 is {\it far} larger
than $1/3$; it is $p({\rm max}_{10}(r) > 4) = 1 - (1 - 1/3)^{10} \approx 0.98$.
The case for the periodogram is analogous: the probability of seeing a
spurious peak at any single location (\eq{LS-prob-z}) is relatively small,
but the probability of seeing a single spurious peak among a large number
of frequencies is much higher.

With the dice, this calculation is easy because the rolls are independent: the
result of one roll does not affect the result of the next.
With the periodogram, though, the value at one frequency is correlated with
the value at other frequencies in a way that is quite difficult to analytically
express---these correlations come from the convolution with the survey
window.
Nonetheless, one common approach to estimating the distribution has been to
assume it can be modeled on some ``effective number'' of independent
frequencies $N_{eff}$, so that the FAP can be estimated as
\begin{equation}
  FAP(z) \approx 1 - \big[P_{single}(z)\big]^{N_{eff}}
  \eqlabel{FAP-neff}
\end{equation}
A very simple estimate for $N_{eff}$ is based on our arguments of the
expected peak width, $\delta f = 1/T$.
In this approximation, the number of independent peaks in a range
$0 \le f \le f_{max}$ is assumed to be $N_{eff} = f_{max} T$.
There have been various attempts to estimate this value more carefully,
both analytically and via simulations
\citep[see, e.g.][]{Horne86,Schwarzenberg-Czerny98,Cumming04,Frescura08},
but all such approaches are necessarily only approximations.

\subsubsubsection{\citet{Baluev2008} Method}

A more sophisticated treatment of the problem is that of \citet{Baluev2008},
who derived an analytic result based on theory of extreme values
for stochastic processes.
For the standard periodogram in \eq{lomb-scargle-periodogram},
\citet{Baluev2008} showed that the following
formula for the FAP provides a close upper-limit even in the case of
highly structured window functions:
\begin{equation}
  FAP(z) \approx 1 - P_{single}(z)e^{-\tau(z)}
  \eqlabel{FAP-baluev}
\end{equation}
where, for the normalized periodogram (\eq{lomb-scargle-normalized}),
\begin{equation}
  \tau(z) \approx W (1 - z)^{(N - 4)/2}\sqrt{z}
\end{equation}
and $W = f_{max} \sqrt{4\pi\ {\rm var}(t)}$ is an effective width of the
observing window in units of the maximum frequency chosen for the analysis.
This result is not an exact measure of the FAP, but rather an
upper limit that is valid for alias-free periodograms, {i.e.} cases
where the window function has very little structure,
but which holds quite well even in the case of more realistic survey windows.

\subsubsubsection{Bootstrap Method}
In the absence of a true analytic solution to the false alarm probability,
we can turn to computational methods such as the bootstrap.
The bootstrap method is a technique in which the statistic in question is
computed repeatedly on many random resamplings of the data in order to
approximate the distribution of that statistic
\citep[see][for a useful general discussion of this technique]{ICVG2014}.
For the periodogram, in each resampling we keep the temporal coordinates
the same, draw observations randomly with replacement from the observed
values, and then compute the maximum of the resulting periodogram.
For enough resamplings, the distribution of these maxima will approximate
the true distribution for the case with no periodic signal present.
The bootstrap produces the most robust estimate of the FAP because it makes
few assumptions about the form of the periodogram distribution,
and fully accounts for survey window effects.

Unfortunately, the computational costs of the bootstrap can be quite
prohibitive: to accurately measure small levels of FAP requires constructing
a large number of full periodograms.
If you are probing a false positive rate of $r$ among $N$ bootstrap reseamples,
you would roughly expect to find $rN \pm \sqrt{rN}$ relevant peaks in your
bootstrap sample.
More concretely, for 1,000 bootstrap samples, you'll find only $\sim 10\pm 3$
peaks reflecting a 1\% false positive rate.
The random fluctuations in that count translate directly to an inability to
accurately estimate the false positive rate at that level.
A good rule-of-thumb is that to accurately characterize a false positive rate
$r$ requires
something like $\sim 10 / r$ individual periodograms to be computed---this grows
computationally intractable quite quickly.
The bootstrap is also not universally applicable: for example, it does not
correctly account for cases where noise in observations is correlated;
for more discussion of the bootstrap approach and its caveats, see
\citet{ICVG2014} and references therein.

\begin{figure}[ht]
  \centering
  \includegraphics[width=\textwidth]{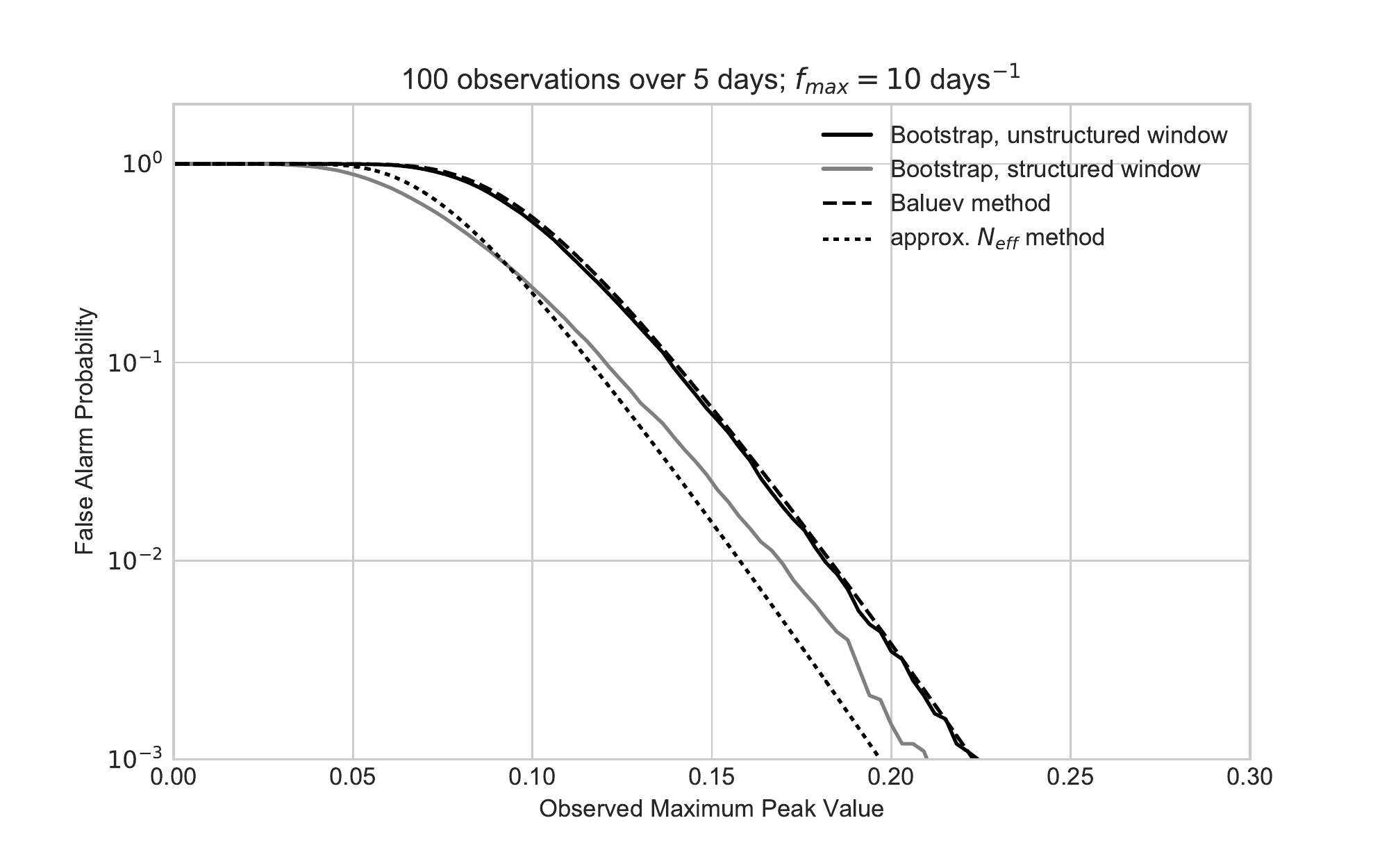}
  \caption{Comparison of various approaches to computing the false alarm
    probability (FAP) for simulated observations with both structured and
    unstructured survey windows. Note that the approximate $N_{eff}$ method
    in its most na{\"i}ve form tends to underestimate the bootstrapped FAP,
    while the \citet{Baluev2008} method tends to overestimate the bootstrapped
    FAP, particularly for data with a highly-structured survey window.
    \figlabel{FAP-bootstrap}}
\end{figure}

\fig{FAP-bootstrap} compares these methods of estimating the false alarm
probability, for both an unstructured survey window, and a structured window
which produces the kinds of aliasing we have discussed above.
The na{\"i}ve approximation in this case under-estimates the FAP at nearly all
levels---so, for example, it might lead you to think a peak has a FAP of 10\%
when in fact it is closer to 30\%.
The Baluev method, by design, over-estimates the FAP, and is quite close to
the bootstraped distribution in the case of an unstructured window.
For a highly structured window, the Baluev method does not perform as well,
but still tends to over-estimate the FAP---so, for example, it might lead you
to think a peak as a FAP of 10\% when in fact it is closer to 5\%.

Finally, I will note that \citet{Suveges12} has shown the promise of a
hybrid approach of the bootstrap method and the extreme value statistics
of \citet{Baluev2008}; this has the ability to compute accurate FAPs
without the need to compute thousands of full bootstrapped periodograms.

\subsubsubsection{What Does the False Alarm Probability Measure?}
\sectlabel{what-FAP-measures}

While the False Alarm Probability can be a useful concept, you should be
careful to keep in mind that it answers a very specific question:
{\it ``what is the probability that a signal with no periodic component would
  lead to a peak of this magnitude?''}.
In particular, it emphatically does {\it not} answer the much more relevant
question {\it ``what is the probability that this dataset is periodic
  given these observations?''.}
This boils down to a case of $P(A \mid B) \ne P(B \mid A)$,
but still it is common to see the false alarm probability
treated as if it speaks to the second rather than the first question.
%In other words, the false alarm probability tells you the false positive rate
%in a frequentist, rather than Bayesian sense: that is, it is a false positive
%rate conditioned on the entire space of possible observations rather than on
%the particular data you observed.

Similarly, the false alarm probability tells us nothing about the false
negative rate; {i.e.} the rate at which
we would expect to miss a periodic signal that is, in fact, present.
It also tells us nothing about the error rate; {i.e.} the rate at which we
would expect to identify an incorrect alias of a signal present in our data
(as we saw in \sect{failure-modes}).
Too often users are tempted to interpret the FAP too broadly, with the hope that
it would speak to questions that are beyond its reach.

Unfortunately, there is no silver bullet for answering
these broader, more relevant questions of uncertainty of Lomb-Scargle results.
Perhaps the most fruitful path toward understanding of such effects for a
particular set of observations---with particular noise characteristics and a
particular observing window---is via simulated data injected into the detection
pipeline.
In fact, this type of simulation is a vital component of most (if not all)
current and future time-domain surveys that seek to
detect, characterize, and catalog periodic objects,
regardless of the particular approach used to identify periodicity
\citep[see, e.g.][etc.]{opsim1, opsim2, Ivezic08LSST, Sesar2010,
  Oluseyi12, McQuillan2012, Drake2014}

\subsection{Periodogram Normalization and Interpretation}
\sectlabel{normalizations}

Often it is useful to be able to interpret the periodogram values themselves.
Recall that there are two equally valid perspectives of what the periodogram
is measuring---the Fourier view and the least-squares view---and each of these
lends itself to a different approach to normalizion and interpretation of the
periodogram.

\subsubsection{PSD normalization}
When considering the periodogram from the Fourier perspective,
it is useful to normalize the periodogram such that in the special case
of equal spaced data, it recovers the standard Fourier power spectrum.
This is the normalization used in \eq{lomb-scargle-periodogram} and the
equivalent least-squares expression seen in \eq{lomb-scargle-chi2}:
\begin{equation}
  P(f) = \frac{1}{2}\big[\hat{\chi}^2_0 - \hat{\chi}^2(f)\big]
  \eqlabel{lomb-scargle-chi2-2}
\end{equation}
For equally-spaced data, this periodogram is identical to the standard
definition based on the fast Fourier transform:
\begin{equation}
  P(f) = \frac{1}{N} \left| FFT(y_n) \right|^2
\end{equation}
in particular, this means that the units of the periodogram are $unit(y)^2$,
and can be interpreted as squared-amplitudes of the Fourier component at
each frequency.
Note, however, that the units change if data uncertainties are incorporated
into the periodogram as in \sect{extensions-observational-noise};
the periodogram in this normalization becomes unitless: it is essentially
a measure of periodic content in signal-to-noise ratio
rather than in signal itself.

\subsubsection{Least-squares normalization}
In the least-squares view of the periodogram, the periodogram is interpreted
as an inverse measure of the goodness-of-fit for a model.
When we express the periodogram as a function of $\chi^2$ model fits as in
\eq{lomb-scargle-chi2-2}, it becomes clear that the periodogram has a
mathematical maximum value: if the sinusoidal model {\it perfectly} fits
the data at some frequency $f_0$, then $\hat{\chi}^2(f_0) = 0$ and the
periodogram is maximized at a value of $\hat{\chi}^2_0 / 2$.
On the other end, it is mathematically impossible for a best-fit sinusoidal
model at any frequency to do worse than the simple constant, non-varying
fit reflected in $\hat{\chi}^2_0$, and so the minimum value of
\eq{lomb-scargle-chi2-2} is exactly zero.

This suggests a different normalization of the periodogram, where the values
are dimensionless and lie between 0 and 1:
\begin{equation}
  P_{norm}(f) = 1 - \frac{\hat{\chi}^2(f)}{\hat{\chi}^2_0}
  \eqlabel{lomb-scargle-normalized}
\end{equation}
This ``normalized periodogram'' is unitless, and
is directly proportional to the unnormalized
(or PSD-normalized) periodogram in \eq{lomb-scargle-chi2-2}.

While the normalization does not affect the shape of the periodogram,
its main practical consequence is that the statistics of the resulting
periodogram differ slightly, and this needs to be taken into account when
computing analytic estimates of uncertainties and false alarm probabilities
explored in \sect{false-alarm-probability}.
Other normalizations exist as well, but seem to be rarely used in practice.
For a concise summary of several periodogram normalizations and their
statistical properties, refer to the introduction of \citet{Baluev2008}.

\subsection{Algorithmic Considerations}
\sectlabel{algorithmic-considerations}
Given the size of the frequency grid required to fully characterize the periods
from a given dataset, it is vital to have an efficient algorithm for evaluating
the periodogram. We will see that the na{\"i}ve implementation of
the standard Lomb-Scargle formula (\eq{lomb-scargle-periodogram})
scales poorly with the size of the data, but that
fast alternatives are available.

Suppose you have a set of $N$ observations over a time-span $T$ for an average
cadence of $\overline{\delta t} = N/T$.
From \eq{n-eval} we see that the number of frequencies we need to evaluate is
$N_f \propto T f_{max} = N f_{max} / \overline{\delta t}$.
Holding constant the average survey cadence $\overline{\delta t}$ and
$f_{max}$, we find that the number of
frequencies required is directly proportional to the number of data points.
The computation of the Lomb-scargle periodogram in \eq{lomb-scargle-periodogram}
requires sums over $N$ sinusoids at each of the
$N_f$ frequencies, and thus we see that the
na{\"i}ve scaling of the algorithm with the size of the dataset is
$\mathcal{O}(N^2)$, when survey properties are held constant.
Due to the trigonometric functions involved, this turns out to be a rather
``expensive'' $\mathcal{O}(N^2)$, which makes the direct
implementation impractical for even modestly-sized datasets.

Fortunately, several faster implementations have been proposed to compute the
periodogram to arbitrary precision in $\mathcal{O}(N\log N)$ time.
The first of these is discussed in \citet{Press89}, which uses an inverse
interpolation operation along with a Fast Fourier Transform to compute the
trigonometric components of \eq{lomb-scargle-periodogram} very efficiently over
a large number of frequencies.
\citet{Zechmeister09} showed how this approach can be easily extended to
the floating mean periodogram discussed in \sect{floating-mean}.

A qualitatively similar approach to the \citet{Press89} algorithm
is presented by \citet{Leroy2012}:
It makes use of the Non-equispaced Fast Fourier Transform
\citep[NFFT, see][]{Keiner2009} to compute the Lomb-Scargle periodogram about
a factor of 10 faster than the \citet{Press89} approach.
For the multi-term models discussed in \sect{multiterm}, \citet{Palmer09}
presents an adaptation of the \citet{Press89} method that can compute the
multi-term result in $\mathcal{O}(NK\log N)$, for $N$ data points and
$K$ Fourier components.

\section{Conclusion and Summary}
\sectlabel{conclusion}

This paper has been a conceptual tour of the Lomb-Scargle periodogram, from
its roots in Fourier analysis, to its equivalence with special cases of
periodic analysis based on least squares model fitting and Bayesian
analysis.
From this conceptual understanding, we considered a list of challenges
and issues to be considered when applying the method in practice.
We will finish here with a brief summary of these practical
recommendations, along with a somewhat opinionated
post-script for further thought.

\subsection{Summary of Recommendations}
\sectlabel{summary}
The previous pages contain a large amount of background and advice for working
with the Lomb-Scargle periodogram.
Following is a brief summary of the considerations to keep in mind when you
apply this algorithm to a dataset:
\begin{enumerate}
  \item Choose an appropriate frequency grid: the minimum can be set to zero,
    the maximum set based on the precision of the time measurements
    (\sect{non-uniform-nyquist}), and the grid spacing set based on the
    temporal baseline of the data (\sect{frequency-grid})
    so as to not sample too coarsely around peaks.
    If this grid size is computationally intractable, reduce the maximum
    frequency based on what kinds of signals you are looking for.
  \item Compute the window transform using the Lomb-Scargle periodogram,
    by substituting $g_n=1$ for each $t_n$ and making sure not to pre-center
    the data or use a floating-mean model (\sect{computing-window-function}).
    Examine this window function for dominant features, such as daily or
    annual aliases (cf. \fig{LINEAR-window})
    or Nyquist-like limits (cf. \fig{kepler-data}).
  \item Compute the periodogram for your data. You should always use the
    floating-mean model (\sect{floating-mean}), as it produces more robust
    periodograms and has few if any disadvantages. Avoid multi-term Fourier
    extensions (\sect{multiterm}) when the signal is unknown, because its
    main effect is to increase periodogram noise
    (cf. \figs{kepler-multiterm}{LINEAR-multiterm}).
  \item Plot the periodogram, and identify any patterns that may be caused
    by features you observed in the window function power. Plot reference
    lines showing several False Alarm Probability levels to understand whether
    your periodogram peaks are significant enough to be labeled detections:
    use the Baluev method or the bootstrap method if it is computationally
    feasible (\sect{false-alarm-probability}).
    Keep in mind exactly what the False Alarm Probability measures, and
    avoid the temptation to misinterpret it (\sect{what-FAP-measures}).
  \item If the window function shows strong aliasing, locate the expected
    multiple maxima and plot the phased light curve at each.
    If there is indication that the sinusoidal model under-fits the data
    (cf. \fig{binary-multiterm}) then consider
    re-fitting with a multi-term Fourier model (\sect{multiterm}).
  \item If you have prior knowledge of the shape of light curves you are
    trying to detect, consider using more complex models to choose between
    multiple peaks in the periodogram (\sect{bayesian-periodograms}).
    This type of refinement can be quite useful in building automated
    pipelines for period fitting, especially in cases where the window
    aliasing is strong.
  \item If you are building an automated pipeline based on Lomb-Scargle for
    use in a survey, consider injecting known signals into the pipeline to
    measure your detection efficiency as a function of object type, brightness,
    and other relevant characteristics (\sect{failure-modes})
\end{enumerate}
This list is certainly not comprehensive for all uses of the periodogram, but
it should serve as a brief reminder of the kinds of issues you should keep
in mind when using the method to detect periodic signals.

\subsection{Postscript: Why Lomb-Scargle?}
\sectlabel{postscript}
After considering all of these practical aspects of the periodogram,
I think it is worth stepping back to
revisit the question of {\it why} astronomers tend to gravitate toward the
Lomb-Scargle approach rather than the (in many ways simpler) classical
periodogram.

As discussed in \sect{schuster-to-lomb-scargle}, the Lomb-Scargle approach has
two distinct benefits over the classical periodogram:
the noise distribution at each individual frequency is chi-square distributed
under the null hypothesis,
and the result is equivalent to a periodogram derived
from a least squares analysis.
But somehow along the way, a mythology seems to have developed surrounding
the efficiency and efficacy of the Lomb-Scargle approach. In particular,
it's common to see statements or implications along the lines of the following:
\begin{itemize}
  \item {\it Myth: The Lomb-Scargle periodogram can be computed more efficiently
    than the classical periodogram.} Reality: computationally, the two
    are quite similar, and in fact the fastest Lomb-Scargle algorithm currently
    available is based on the classical periodogram computed via the
    the NFFT algorithm (see \sect{algorithmic-considerations}).
  \item {\it Myth: The Lomb-Scargle periodogram is faster than a direct
    least squares periodogram because it avoids explicitly solving for
    model coefficients.}
    Reality: model coefficients can be determined with little extra
    computation \citep[see the discussion in][]{ICVG2014}.
  \item {\it Myth: The Lomb-Scargle periodogram allows analytic computation of
    statistics for periodogram peaks.} Reality: while this is true at
    individual frequencies, it is not true for the more relevant question of
    maximum peak heights across multiple frequencies,
    which must be either approximated
    or computed by bootstrap resampling (see \sect{uncertainties})
  \item {\it Myth: The Lomb-Scargle periodogram corrects for aliasing due to
    sampling and leads to independent noise at each frequency.}
    Reality: for structured window functions common to most astronomical
    applications, the Lomb-Scargle
    periodogram has the same window-driven issues as the classical
    periodogram, including spurious peaks due to partial aliasing, and
    highly correlated periodogram errors (see \sect{failure-modes}).
  \item {\it Myth: Bayesian analysis shows that Lomb-Scargle is the optimal
    statistic for detecting periodic signals in data.} Reality: Bayesian
    analysis shows that Lomb-Scargle is the optimal statistic for
    {\it fitting a sinusoid to data}, which is not the same as saying it
    is optimal for finding the frequency of a generic, potentially
    non-sinusoidal signal (see \sect{bayesian-periodograms}).
\end{itemize}
With these misconceptions corrected, what is the practical advantage of
Lomb-Scargle over a classical periodogram?
What would we lose if we instead used the simple classical Fourier periodogram,
estimating uncertainty, significance, and false alarm probability by
resampling and simulation, as we must for Lomb-Scargle itself?

The advantage of analytic statistics for Lomb-Scargle evaporates in light of the
need to account for multiple frequencies, so the only advantage left is the
correspondence to least squares and Bayesian models, and in particular the
ability to generalize to more complicated models where appropriate---but in
this case you're not really using Lomb-Scargle at all,
but rather a generative Bayesian model for your
data based on some strong prior information about the form of your signal.
The equivalence of Lomb-Scargle to a Bayesian sinusoidal model is perhaps
an interesting bit of trivia, but not itself a reason to use that model if your
data is not known {\it a priori} to be sinusoidal---it could even be construed as
an argument {\it against} Lomb-Scargle in the general case where the assumption
of a sinusoid is not well-founded.

Conversely, if you replace your Lomb-Scargle approach with a classical periodogram,
what you gain is the ability to reason quantitatively about the effects of the
survey window function on the resulting periodogram (cf. \sect{CLEAN}).
While the deconvolution problem is ill-posed, there is no reason to assume
this is a fatal defect: ill-posed linear models are solved routinely in
other areas of computational research, particularly by using sparsity
priors or various forms of regularization.
In any case, I would contend that there is ample room for practitioners to
question the prevailing folk wisdom of the advantage of Lomb-Scargle
over approaches based directly on the Fourier transform and classical
periodogram.

\subsection{Figures and Code}

All computations and figures in this paper were produced using Python,
and in particular used the
Numpy \citep{numpy, numpybook},
Pandas \citep{pandas},
AstroPy \citep{Astropy2013},
and Matplotlib \citep{matplotlib} packages.
Periodograms where computed using the
AstroPy implementations\footnote{The AstroPy Lomb-Scargle documentation is at
\url{http://docs.astropy.org/en/stable/stats/lombscargle.html}}
of the \citet{Press89}, \citet{Zechmeister09}, and
\citet{Palmer09} algorithms, which were adapted from Python code originally
published by \citet{ICVG2014} and \citet{VanderPlas2015}.
All code behind this paper, including code to reproduce all figures,
is available in the form of Jupyter notebooks in the paper's GitHub
repository\footnote{Source code \& notebooks available at
\url{http://github.com/jakevdp/PracticalLombScargle/}}.

\acknowledgments
{\it Acknowledgments:} I am grateful to Jeff Scargle, Max Mahlke, Dan
Foreman-Mackey, Zeljko Ivezic, and David Hogg for helpful feedback
on early drafts of this paper.
This work was supported by the University of Washington eScience Institute,
with funding from the Gordon and Betty Moore Foundation, the Alfred P. Sloan
Foundation, and the Washington Research Foundation.

\bibliographystyle{apj}
\bibliography{PracticalLombScargle}

\end{document}